%% file: Bitcoin_paper_-_19092021.tex
\documentclass[11pt,a4paper]{article}

\usepackage[utf8]{inputenc}
\usepackage[english]{babel}
\usepackage{amsmath,amsfonts,dsfont,amssymb,graphicx,rotating, setspace, booktabs, tikz, natbib, xcolor, enumerate, array, titling, dcolumn}
\usepackage[toc,page]{appendix}
\usepackage[left=2cm,right=2cm,top=3cm,bottom=3cm]{geometry}
\usepackage[skip=10pt, font=small, labelfont=bf]{caption}
\usepackage{hyperref}
\hypersetup{
    colorlinks=true,
    linkcolor=blue,
    citecolor=blue,
    filecolor=blue,      
    urlcolor=cyan,
}

\graphicspath{{Images/}}
\newtheorem{defn}{Definition}
\newtheorem{assumpt}{Assumption}
\newtheorem{example}{Example}
\newtheorem{theorem}{Theorem}

\newcommand{\info}[2]{\mathcal{#1}_{#2}}

\newcommand{\bs}[1]{\boldsymbol{#1}}

\DeclareMathOperator{\w}{w}
\DeclareMathOperator{\W}{W}
\DeclareMathOperator{\Y}{Y}
\DeclareMathOperator{\X}{X}
\DeclareMathOperator{\V}{V}

\author{\Large Fiammetta Menchetti \\ DiSIA, University of Florence, \texttt{fiammetta.menchetti@unifi.it} 
\\ \Large Fabrizio Cipollini \\ DiSIA, University of Florence,  \texttt{fabrizio.cipollini@unifi.it} 
\\ \Large Fabrizia Mealli \\ DiSIA, University of Florence,  \texttt{fabrizia.mealli@unifi.it}}
\title{\huge \textbf{Causal effect of regulated Bitcoin futures \\ on volatility and volume}}
\date{\today}

\begin{document}
\maketitle

\begin{abstract}
In December 2017, two leading derivative exchanges, CBOE and CME, introduced the first regulated Bitcoin futures. Our aim is estimating their causal impact on Bitcoin volatility and trading volume. Employing a new causal approach, C-ARIMA, we find that the CME future triggered an increase in both outcomes. There is also evidence of a positive volume-volatility relationship and that the effect on volatility was partially due to the higher trading volumes induced by the launch of the contract. After controlling for the effect on volumes, we find that the CME instrument caused Bitcoin volatility to increase by more than double. 
\end{abstract}

\bigskip
\bigskip
\textit{Keywords:} Bitcoin, Causal inference, Counterfactual analysis, Futures, Volatility, Volume 

\newpage

\section{Introduction}
\label{sect:intro}
The attitude toward cryptocurrencies has radically changed in recent years: the initial skepticism that once accompanied them has given way to enthusiastic pronouncements, as their market capitalization reaches new record levels. Nonetheless, Bitcoin share of the entire cryptocurrency capitalization is above 40\% and concern is rising that it may become the ultimate speculative asset class, as people and financial institutions get increasingly involved with it.\footnote{According to \citet{coinmarketcap}, at the time of writing (September 2021) the global crypto market cap amounts to \$2.11 trillion, more than the current GDP of countries such as Italy and Canada \citep{worldbank}. For a detailed analysis on the growing interactions between Bitcoin and the financial sector, see \citet{bloombergcryptos}.}

A turning point in Bitcoin history could have been the launch of the first two regulated Bitcoin futures
by the Chigago Board of Exchange (CBOE) and the Chicago Mercantile Exchange (CME) in December 2017. Before that time, Bitcoin derivatives were only traded over-the-counter but since the introduction of a regulated derivative market, the interest of the financial sector toward Bitcoin has surged: hedge funds are investing in Bitcoin and banks are offering safekeeping services to their clients \citep{Leask:2020, Ossinger:2021, Sullivan:2021}.

Whether this is enhancing price discovery or increasing volatility of an already hyper-volatile asset is still unclear and the literature is far from a consensus. \citet{Hafner:2020} find evidence of a bubble-like behavior in Bitcoin returns and growing volatility in late 2017, suggesting it might be due to the new futures contracts and the increased media coverage during this period. Conversely, \citet{Shi:2017} finds a significant decrease in the spot volatility after the launch of CBOE contract, whereas later studies show that Bitcoin volatility increased right after the launch of CME future \citep{Kim:Lee:Kang:2020} as well as around its announcement date \citep{Corbet:Lucey:Peat:Vigne:2018}. Interestingly, the peak reached by Bitcoin price on December 16, 2017 matched the launch of the CME future; since then and for over a year, prices have experienced a sharp decline, suggesting that the newly introduced instruments may have allowed the entrance of ``pessimistic’’ traders willing to bet for a price drop but unable to do so until the creation of a derivative market \citep{Hale:Krishnamurthy:Kudlyak:Shultz:2018}. 

Albeit the relatively few studies on the effect of Bitcoin futures, the impact of derivatives trading on the underlying spot volatility has been thoroughly investigated for instruments such as stocks and financial indexes. In particular, there are two conflicting theories about the effect of futures on the underlying spot markets and empirical evidence is mixed: some warn that speculative behaviors and information asymmetries brought by futures markets may increase volatility and jeopardize the value formation process in the spot market \citep{Stein:1987,Figlewski:1981,Antoniou:Holmes:1995,Harris:1989}; others argue that futures trading enhances information flows and price discovery, thereby reducing volatility and stabilizing the underlying spot market prices \citep{Danthine:1978,Moriarty:Tosini:1985,Edwards:1988,Bessembinder:Seguin:1992,Antoniou:Holmes:Priestley:1998,Jochum:Kondres:1998}. Recent works on individual stock futures \citep{Mckenzie:Brailsford:Faff:2001} and emerging economies \citep{Baklaci:Tutek:2006, Chen:Han:Li:2013,Bohl:Diesteldorf:Siklos:2015} support the idea that futures trading acts as a stabilizing force.

Contributing to this debate, in this paper we estimate the causal effect of both the CBOE and the CME instruments on Bitcoin volatility; we also investigate the presence of possible announcement effects and the causal impact on Bitcoin trading volume.


Indeed, the positive association between volume and volatility is well documented in the literature (e.g.,\citet{Bessembinder:Seguin:1993, Chen:Firth:Rui:2001, Kim:2005, Gebka:Wohar:2013}); therefore, part of the positive (negative) effect that we might observe on Bitcoin volatility could be the result of an indirect impact exerted by increased (decreased) trading volume. To verify this claim, we also estimate the effect of Bitcoin futures on trading volume and we explore the presence of a possible connection to volatility. Recent studies have attempted to investigate whether the positive relation between volume and volatility is present in the Bitcoin market as well, but the evidence is still mixed. Using squared returns as a proxy of volatility and a methodology based on non parametric causality-in-quantiles test, \citet{Balcilar:Bouri:Gupta:Roubaud:2017} find that trading volumes cannot predict the volatility of Bitcoin returns at any point of the conditional distribution. However, opposite results are obtained by the authors when repeating the analysis using GARCH volatility. Similarly, \citet{Bouri:Lau:Lucey:Roubaud:2019} show that trading volume do not Granger-cause Bitcoin volatility as proxied by squared returns. Conversely, using realized volatility computed from high-frequency data, \citet{Aalborg:Molnar:deVries:2019} find that volumes Granger cause volatility.


From the methodological viewpoint, most of the aforementioned studies employ Granger causality and only \citet{Kim:Lee:Kang:2020} uses a causal approach based on counterfactual outcomes --- Difference-in-Difference (DiD)--- in the attempt to attribute the uncovered effect to the introduction of futures.
Nonetheless, DiD relies on the assumption that the treated unit in the absence of intervention would have experienced the same trend as the control unit; therefore, this approach is impractical in all situations where, as in our case, there is no reliable control group available or the parallel trend assumption is troubled \citep{ONeill:Kreif:Grieve:Sutton:Sekhon:2016}.

In this paper, we employ the novel C-ARIMA approach proposed in \citet{Menchetti:Cipollini:Mealli:2021}, which, in contrast to DiD, allows to estimate causal effects in time series settings where suitable control series are unavailable and without imposing parallel trends. Furthermore, unlike traditional intervention analysis \citep{Box:Tiao:1975,Box:Tiao:1976, Bhattacharyya:Layton:1979, Larcker:Gordon:1980, Balke:Fomby:1994}, this method enables the computation of properly defined ``causal'' effects. 
Indeed, C-ARIMA is based on the Rubin's potential outcomes approach \citep{Rubin:1974, Rubin:1975, Rubin:1978, Imbens:Rubin:2015} that allows to define the effect of an intervention as a contrast of ``potential outcomes'' as well as to discuss assumptions enabling the attribution of the effect to the intervention. Following a causal framework clarifying the assumptions needed to define and estimate the effect from available data, C-ARIMA uses ARIMA models to estimate the causal impact of the intervention by contrasting the observed post-intervention series with the counterfactual series that we would have had in the absence of intervention.

Although it was developed in a setting where a single intervention occurs, in this paper we extend the C-ARIMA approach to estimate causal effects in a multi-intervention setting, namely, a situation where a time series is subject to multiple interventions over time. 
We then use it to evaluate the impact of the first two regulated Bitcoin futures on Bitcoin volatility as proxied by the Garman-Klass estimator and on transaction volumes. Our results indicate that CME future triggered an increase in both outcomes; there is also evidence that the positive association between volumes and volatility holds in the Bitcoin market as well and that the higher volumes induced by the CME instrument contributed, at least in part, to the observed effect on volatility. By controlling for these higher volumes, we find that Bitcoin volatility more than doubled due to the CME contract. Conversely, the CBOE instrument had only a minor impact on volumes and we do not find evidence of announcement effects.
Overall, our results seem to support the idea that futures trading is driving Bitcoin volatility upward, instead of diminishing it; this might be due to a renovated interest among investors following a perception of greater transparency, or to increased speculation, since regulated Bitcoin futures allow a greater number of people to bet against Bitcoin price. Regulators should be aware of these implications as banks and financial institutions are opening up to this new investment opportunity.

The rest of the paper is structured as follows: Section \ref{sect:background} describes the background for our empirical analysis; Section \ref{sect:causal_framework} sets out the causal framework; Section \ref{sect:cARIMA} extends the C-ARIMA model to the multi-intervention setting; Sections \ref{sect:data_method} and \ref{sect:results} illustrate the data and the results of the empirical analysis; Section \ref{sect:conclusions} concludes.

\section{Background}
\label{sect:background}

Bitcoin is a peer-to-peer payment system created in 2009 
under the pseudonym of Satoshi Nakamoto \citep{Nakamoto:2008}. In contrast to fiat money relying on central banks and intermediaries, Bitcoin is decentralized, meaning that its value is not backed by any central bank. 
In short, transaction data are recorded in \textit{blocks}, each containing a reference (the hash) to the previous ones, thereby forming a chain know as the \textit{blockchain}. Transactions are validated through the so-called \textit{mining} process, which requires to find the hash by solving a time consuming cryptographic problem. Every time this happens, the new validated block is added to the chain and miners are rewarded with new Bitcoins. Thus, essentially, the blockchain constitutes an electronic public ledger of validated transactions, which is stored and updated on miner's computers (the network nodes) to avoid double spending problems and frauds. Mining is the only way new Bitcoins are introduced in the system and their supply is limited by design to a maximum of 21 million units that will be reached by 2140.\footnote{More precisely, an algorithm adjusts the difficulty of the numerical problem for finding the hash based on the network performance, so that new blocks are generated at a fixed rate (every 10 minutes); in addition, the number of generated Bitcoins is halved by design every 210,000 blocks (approximately 4 years). Considering these facts, it can be estimated that by 2140 miners' reward will be roughly zero, meaning that the maximum supply will be reached.} 

Bitcoins are traded on multiple exchanges, the top A-rated ones being Coinbase, Gemini and Bitstamp \citep{CryptoBenchmark:2021}\footnote{In the Exchange Benchmark Report released in February 2021, CryptoCompare rated 162 exchanges according to 68 qualitative and quantitative metrics. They show that existing metrics, such as volume or liquidity, can be easily manipulated (e.g. volumes can be inflated through strategies such as trading competitions, airdrops and transaction fee mining) and thus are inadequate to reflect the reliability of the trading venue. For further details regarding these strategies and the rating methodology see \citet{CryptoBenchmark:2021}.}. At the time of writing (September 2021, timestamp 2021-09-16 00:00:00 GMT) Bitcoin is trading at \$47,629.49 while in 2010 1BTC was valued \$0.05.
Such extreme price fluctuations (and the related opportunities of huge profits) probably contributed to spark investor interest toward cryptocurrencies and  
to the decision, by two major derivative exchanges, of starting a regulated derivative market for Bitcoin futures. Specifically, on August 2, 2017 CBOE announced its partnership with
Gemini Trust Company to use Gemini’s Bitcoin market data in the creation of derivatives products for trading; the future was then released on December 10. Meanwhile, CME  announced the launch of its contract becoming effective starting from December 18. 
Since then, having seen its market share quickly eroded by the new future \citep{Baydakova:2019}, CBOE has stopped listing additional Bitcoin futures for trading and after nine months from the release of the first contract, also the Intercontinental Exchange (ICE) started offering Bitcoin derivatives. Table \ref{tab:event_history} summarizes the major events occurring in Bitcoin futures' history up to December 2017. 

In our empirical analysis, we focus on the futures launched by CBOE and CME, as it is commonly done in the literature, to assess whether regulated futures have a role in determining Bitcoin volatility and volume.  

To improve the prediction accuracy of our models, we include covariates that might be linked to Bitcoin trading. 
Early works evidence association between Bitcoin price and variables such as the exchange-trade ratio (i.e. ratio between trade volume and transaction volume), hash rate, Google and Wikipedia queries and the Shangai index \citep{Kristoufek:2013,Kristoufek:2015,Bouoiyour:Selmi:2015}. In a later study,  \citet{Xin:Wang:2017} find a strong association between Bitcoin price and key economic fundamentals (US interest rate, USD money supply, Bitcoin money supply and transaction volume); they also find a connection between technology factors such as mining difficulty and public recognition and the price before 2014.\footnote{A major event that affected the Bitcoin network is the failure of Mt. Gox, the leading Bitcoin exchange until February 28, 2014, when it filed for bankruptcy protection after announcing a theft of about 850,000 BTC following a security breach \citep{McLannahan:2014,Cermak:2017}. This is considered in the literature as a path-breaking event and many works focusing on Bitcoin returns make a distinction between an early period (before Mt. Gox failure) and a later period (after the failure).} Similarly, \citet{Ciaian:Kancs:Rajcaniova:2016} evidence that market forces of Bitcoin supply and demand are strongly related to Bitcoin price changes while proxies of public recognition (i.e. views on Wikipedia, new posts and new members on Bitcoin forums) are mainly associated with the price before 2014. Finally, \citet{Liu:Tsyvinski:2021} find that network factors such as the number of wallet users and the number of active addresses, are important drivers of cryptocurrency price. In Section \ref{sect:data_method} we describe how to include these drivers in our models. 

\begin{table}[!h]
\centering
\caption{Major events in Bitcoin futures' history up to December 2017}
\label{tab:event_history}
\begin{tabular}{ll} \toprule
Date & Event \\ \midrule
2017/08/02 & CBOE announces launch of futures by 2017-Q4 \\
2017/10/31 & CME announces launch of futures by 2017-Q4\\
2017/12/01 & CME announces launch of futures on Dec 18 \\
2017/12/04 & CBOE announces launch of futures on Dec 10 \\
2017/12/10 & Launch of CBOE futures \\
2017/12/18 & Launch of CME futures \\
\bottomrule
\end{tabular}
\end{table}


\section{Causal Framework}
\label{sect:causal_framework}

In this section we illustrate the framework that enables us to make causal statements in a multi-intervention time series setting where the treatments are not randomized.  We first discuss the assumptions on the treatments, potential outcomes and covariates, then we define the causal estimands. In doing so, we refer to the causal framework that is commonly adopted in observational settings with a single intervention \citep{ Choirat:Dominici:Mealli:Papadogeorgou:Wasfy:Zigler:2018, Menchetti:Bojinov:2021,Menchetti:Cipollini:Mealli:2021} and, when needed, we integrate it in order to encompass the multi-intervention situation.    

\subsection{Assumptions}
\label{subsect:assumptions}

For a generic statistical unit, let $\W_t \in \{ 0,1 \}$ be the treatment assignment at time $t \in \{1,\dots, T\}$, where $\{1\}$ denotes that a ``treatment'' (or ``intervention'') has taken place and $\{0\}$ denotes control. Then, $\W_{1:T} = (\W_1,\dots,\W_T) $ is the assignment path, i.e., the sequence of treatments received by the unit. A realization of $\W_t$ is denoted with the lower case letter $\w_t$.

In our empirical application, the statistical unit is Bitcoin cryptocurrency and the main goal of the analysis is investigating the effect on both volatility and volume generated by the launch of Bitcoin futures by two major regulated exchanges. Since both exchanges disclosed their plans to develop Bitcoin futures, we have two types of interventions: i) the announcements about the upcoming futures; ii) the actual introduction of the two contracts. To make reliable estimates of the effects attributable to these interventions, in accordance with the causal framework presented in \citet{Menchetti:Cipollini:Mealli:2021} we need to discuss three assumptions: non-anticipating potential outcomes, covariates-treatment independence and non-anticipating treatments.

When a statistical unit is assigned to treatment, its outcome may differ from the one we would have observed if the same unit had been assigned to control. In the Rubin causal framework these are called ``potential outcomes''. Let $\Y_t (\w_t)$ denote the potential outcome at time $t$ for a treatment $\w_t$: \textit{non-anticipating potential outcomes} can be function of past treatments but they are unaffected by future treatments and the dependency on the past treatment sequence is denoted by $\Y_t(\w_{1:t})$. In our application, we admit that today volatility is the result of previous announcements and of the trading activity of existing futures but we exclude any influence arising from future announcements and future contracts. In other words, we are ruling out the possibility that market participants have access to privileged information.

In case of binary treatments, like the ones we are considering, at each point in time there are two potential outcomes corresponding, respectively, to the outcome in the presence and in the absence of the intervention, i.e., $\Y_t(1)$ and $\Y_t(0)$. When the intervention occurs, its causal effect is given by direct comparison between them; however, only $\Y_t(1)$ is actually observed, whereas $\Y_t(0)$ is commonly referred as the ``missing'' or ``counterfactual'' outcome and needs to be estimated. In this process, informative covariates may help. 

As outlined in Section \ref{sect:background}, Bitcoin price might be related to several economic and technology factors, such as USD money supply and mining difficulty. Including their values in our analysis may enhance the prediction accuracy of the counterfactual outcome and, as a result, the reliability of the estimated causal effect. However, we should exclude from our covariates set any predictor that might be influenced by the intervention, otherwise the estimated effect will be biased. This \textit{covariates-treatment independence} assumption also applies in our multi-intervention setting; therefore, in the empirical application we use as predictors only those covariates for which such assumption is plausible. For example, since the mining process is ruled by an algorithm and new Bitcoins are created at a fixed rate, the current Bitcoin supply is unaffected by the introduction of futures; conversely, network factors such as the number of wallets and all the proxies of investors' interests (e.g., Google and Wikipedia queries) may have increased following the introduction of the new futures and thus can not be assumed independent of the treatment.  

We also assume \textit{non-anticipating treatments}, meaning that the assignment mechanism (i.e., the process that determines the units receiving treatment) depends solely on past outcomes and past covariates. To understand the importance of such an assumption, consider as an example a research on a new drug where doctors are asked to select the participants among their patients. If doctors assign to treatment only those patients they believe have better chance to complete the treatment successfully and without side effects, the results of such study would be biased due to a ``confounded'' assignment to treatment. A non-anticipating treatment in a time series setting is the analogous of the unconfounded assignment mechanism in the cross-sectional setting \citep{Imbens:Rubin:2015}, since it 
is essential in ensuring that, conditioning on past outcomes and covariates, any difference in the potential outcomes is due to the treatment. As a result, in our application, the effect of the CBOE future is correctly defined up to one week after its launch (beyond that point, the effect would be confounded by the introduction of the CME contract); meanwhile, this assumption ensures that the effect of the CME contract is not confounded by the first future, since we condition on past events.

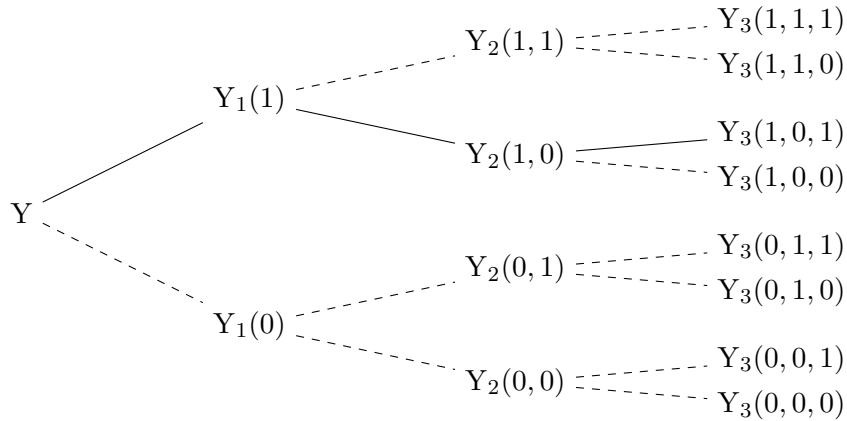
\begin{figure}[!h]
\centering
\caption{Potential outcome time series when $T=3$ and two treatments are administered at times $t_1 = 1$ and $t_2=3$. The total number of potential outcomes is $2(2^3-1) = 14$ corresponding to $2^3$ paths. Only one path is actually observed (indicated by the solid line) whereas the others are missing (indicated by the dashed lines).}
\label{fig:potential_paths}
\begin{tikzpicture}
\tikzstyle{level 1}=[level distance=3cm, sibling distance=3cm]
\tikzstyle{level 2}=[level distance=3.5cm, sibling distance=1.5cm]
\tikzstyle{level 3}=[level distance=3.5cm, sibling distance=0.6cm]
\node{Y}[grow=right]
 child[dashed]{node{$\Y_1(0)$}
  child{node{$\Y_2(0,0)$}
   child{node{$\Y_3(0,0,0)$}}
   child{node{$\Y_3(0,0,1)$}}  
  }
  child{node{$\Y_2(0,1)$}
   child{node{$\Y_3(0,1,0)$}}
   child{node{$\Y_3(0,1,1)$}}  
  }
 }
 child{node{$\Y_1(1)$} edge from parent[black]
  child[black]{node{$\Y_2(1,0)$} edge from parent[black]
   child[dashed]{node{$\Y_3(1,0,0)$}}
   child{node{$\Y_3(1,0,1)$}}  
  }
  child[dashed]{node{$\Y_2(1,1)$} 
   child[black]{node{$\Y_3(1,1,0)$}edge from parent[black]}
   child{node{$\Y_3(1,1,1)$}} 
  }
 };
\end{tikzpicture}
\end{figure}

In a time series setting, interventions can occur at every time point, like the announcements made by the two exchanges. In this case, the total number of potential outcomes at the end of the analysis period is $2(2^T-1)$, corresponding to $2^T$ potential paths (Figure \ref{fig:potential_paths} provides an illustration in a simple multi-intervention setting). However, it is not uncommon to observe a single intervention affecting the time series in a persistent way, such as a price discount for several weeks in a row during a marketing campaign \citep{Brodersen:Gallusser:Koehler:Remy:Scott:2015, Menchetti:Cipollini:Mealli:2021, Menchetti:Bojinov:2021}. As in the case of the two futures, we may also have multiple persistent interventions during the analysis period.

\begin{assumpt}[\textit{N} Persistent interventions]
\label{assumpt:n_persistent_int}
Indicating with $\Lambda = \{t_1,\dots,t_N \}$ the subset of time points at which the interventions take place, we say that the unit received $N$ persistent interventions, if for all $t \le t_1$ we have $\W_{t}=(0,\dots,0)$ and for all $t,t' \in \{ t_{n},\dots,t_{n+1}-1\}$ we have $\W_{t}=\W_{t'}$; we also assume that the effects generated by the N persistent interventions are additive. 
\end{assumpt}

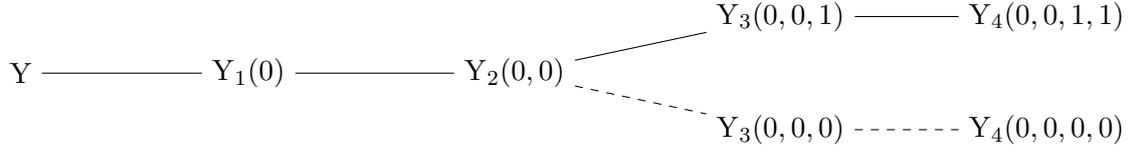
\begin{figure}[h!]
\centering
\caption{Potential outcome time series of a single persistent intervention occurring at time $t = 3$. The solid line represents the observed series while the dashed line depicts the missing potential outcome series.}
\label{fig:potential_paths_persistent}
\begin{tikzpicture}
\tikzstyle{level 1}=[level distance=3cm, sibling distance=3cm]
\tikzstyle{level 2}=[level distance=3.5cm, sibling distance=1.5cm]
\tikzstyle{level 3}=[level distance=3.5cm, sibling distance=1.5cm]
\node{Y}[grow=right]
 child{node{$\Y_1(0)$}
  child{node{$\Y_2(0,0)$}
   child[dashed]{node{$\Y_3(0,0,0)$}
    child[dashed]{node{$\Y_4(0,0,0,0)$}}}
   child{node{$\Y_3(0,0,1)$}
    child{node{$\Y_4(0,0,1,1)$}}}  
  }  
 }
 ;
\end{tikzpicture}
\end{figure}

In words, at each time point in the interval between two subsequent interventions, the statistical unit is either always treated or always assigned to control. Additivity of the effects follows quite naturally: if the interventions persist over time, their effect will likely do the same. Therefore, the observed outcome is the result of all previous interventions.\footnote{Notice that the assumption of additive effects does not pose any strong restriction: in most cases, we can still recover an additive structure by applying simple transformations (e.g., logarithmic transformation).}
In our empirical application, the two futures qualify as persistent interventions, since they are bound to affect permanently the dynamics of Bitcoin price. Indeed, even though the exchange can withdraw its future from the market at any time, the future would trade until its expiration date, which is standardized and set up in advance. 
In this special setting, Assumption \ref{assumpt:n_persistent_int} narrows down the set of potential paths: instead of $2^T$ potential paths, in the time interval between two persistent interventions, $ t \in \{t_{n},\dots, t_{n+1}-1 \}$ we only have two possible paths, 
$$\w_{t_{n}:t_{n+1}-1} = (\underbrace{1,\dots,1}_{t \in \{t_n,\dots,t_{n+1}-1\}})\hspace{8pt}
\text{and} \hspace{8pt} \w'_{t_{n}:t_{n+1}-1} = (\underbrace{0,\dots,0}_{t \in \{t_n,\dots,t_{n+1}-1\}})$$
where $\w_{t_{n}:t_{n+1}-1} = \bs{1}$ is a vector of the observed treatment path indicating that the $n$-th persistent intervention has occurred and persists over time, $\w'_{t_{n}:t_{n+1}-1} = \bs{0}$ is the counterfactual treatment path. Figure \ref{fig:potential_paths_persistent} gives an example of the potential outcomes and treatment paths in a simple case of a single persistent intervention.

\subsection{Causal Estimands}
\label{subsect:estimands}

We can now define the causal estimands of interest. In particular, we provide definitions for two causal effects: the contemporaneous and the pointwise causal effects. The former is used to describe the effect of the announcements, which is reasonably short-term, whereas the latter is used to describe the impact produced by the futures, which qualify as persistent interventions. 

\begin{defn}[Contemporaneous effects]
Indicating with $\Lambda = \{t_1, \dots, t_N \}$ the subset of time points at which the active treatment is administered, the contemporaneous causal effect of the $n$-th treatment at time $t_n \in \Lambda$ conditioning on the observed treatment path $\w_{1:t_n-1}^{obs}$ is,
\begin{equation}
\label{eqn:cont}
\tau^{(n)}(\w_{1:t_n-1}^{obs}, 1;0) = \Y_{t_n}(\w_{1:t_n-1}^{obs},1) - \Y_{t_n}(\w_{1:t_n-1}^{obs},0) .
\end{equation}
\end{defn}
In words, this is the instant effect of a treatment conditioning on the observed treatment path up to that time point. The contemporaneous effect is similar to the general effect in \citep{Bojinov:Shephard:2019}, with the difference that the latter estimand refers to an experimental setting where the intervention is randomly allocated at any point in time and thus it is not necessary to condition on the treatment path. 

\begin{example}
Assume that the solid line in Figure \ref{fig:potential_paths} represents the observed path of the announcements. At $t_1=1$ we immediately observe an intervention and, since there is no past history, the contemporaneous effect is $\tau^{(1)}(1;0) = \Y_1(1) - \Y_1(0)$. At $t_1+1=2$ we observe control. Then, at $t_2=3$ we have a second announcement and by conditioning on the observed path, we can define its contemporaneous causal effect as $\tau^{(2)}(\w^{obs}_{1:2}, 1;0) = \Y_3(1,0,1) - \Y_3(1,0,0)$. 
\end{example}

We now define a class of causal estimands measuring the effect of the $n$-th persistent intervention. Recall that, conditioning on previous treatments, there are only two possible potential paths in the time interval between two persistent interventions, $\w_{t_n:t_{n+1}-1} = \bs{1}$ and $\w'_{t_n:t_{n+1}-1} = \bs{0}$.  

\begin{defn}[Pointwise effects]
Indicating with $\Lambda = \{t_1, \dots, t_N \}$ the subset of time points at which $N$ persistent interventions take place, the point effect at time $t \in \{t_n,\dots,t_{n+1}-1 \}$ of the $n$-th persistent intervention conditioning on the observed path $\w^{obs}_{1:t_n-1}$ is
\begin{equation}
\label{eqn:point}
\tau_t^{(n)}(\w^{obs}_{1:t_n-1},\bs{1}; \bs{0} ) = \Y_t(\w^{obs}_{1:t_n-1}, \bs{1}) - \Y_t(\w^{obs}_{1:t_n-1}, \bs{0}).
\end{equation}
Thus, the cumulative pointwise effect of the $n$-th intervention up to time $t$ is,
\begin{equation}
\label{eqn:cum_pointwise}
\Delta_t^{(n)}(\w^{obs}_{1:t_n-1},\bs{1}; \bs{0}) = \sum\limits_{s=t_n}^t\tau_s^{(n)}(\w^{obs}_{1:t_n-1},\bs{1}; \bs{0})
\end{equation}
and the temporal average pointwise effect of the $n$-th intervention is,
\begin{equation}
\label{eqn:temp_avg_pointwise}
\bar{\tau}_t^{(n)}(\w^{obs}_{1:t_n-1},\bs{1}; \bs{0}) = \frac{\Delta_t^{(n)}( \w^{obs}_{1:t_n-1},\bs{1}; \bs{0})}{t-t_n + 1} .
\end{equation}
\end{defn}  
Notice that for $t = t_n$, the pointwise effect collapses to the contemporaneous effect and in case of one single persistent intervention ($N = 1$) it matches the pointwise effect as defined in \citet{Menchetti:Cipollini:Mealli:2021}.

\begin{example}
Assume that CBOE introduces Bitcoin futures at time $t_1=3$, as outlined in Figure \ref{fig:potential_paths_persistent}. The point effect at $t_1$ is the contemporaneous effect $\tau^{(1)}(\w^{obs}_{1:2},1;0) = \Y_3(0,0,1) - \Y_3(0,0,0)$ and the point effect at time $t_1+1 = 4$ is $\tau_{4}^{(1)}(\w^{obs}_{1:2},1,1;0,0) = \Y_4(0,0,1,1) - \Y_4(0,0,0,0)$. Then, the cumulative and the temporal average pointwise effects are, respectively, $\Delta_4^{(1)}(\w^{obs}_{1:2},1,1;0,0) = \tau^{(1)}(\w^{obs}_{1:2},1;0) +  \tau_{4}^{(1)}(\w^{obs}_{1:2},1,1;0,0)$ and $\bar{\tau}_4^{(1)}(\w^{obs}_{1:2},1,1;0,0) = \frac{1}{2}\Delta_4^{(1)}(\w^{obs}_{1:2},1,1;0,0)$.
\end{example}

Thus, the contemporaneous effect can also be interpreted as a limiting case of the pointwise effect, namely, the effect of a ``pulse'' intervention that lasts only for one time point. As a result, in the next section, the estimators of the contemporaneous effects will be derived as limiting cases of the pointwise effects. 

 
\section{C-ARIMA}
\label{sect:cARIMA}
In this section, we extend the C-ARIMA model \citep{Menchetti:Cipollini:Mealli:2021} to the multi-intervention setting and we use it to derive estimators for the causal effects defined in Section \ref{subsect:estimands}. 

\subsection{The model}

The general formulation of the C-ARIMA model in a setting with a single intervention occurring at time $t^*$ is a linear regression with seasonal ARIMA errors and the addition of a component $\tau_t$,
\begin{equation}
\label{eqn:arima_general}
(1-L^s)^D (1-L)^d \Y_t(\w)  =  \frac{\Theta_Q (L^s)\theta_q(L) }{ \Phi_P(L^s) \phi_p(L)}\varepsilon_t +  (1-L^s)^D (1-L)^d \X_t' \beta + \tau_t \mathds{1}_{\{\w = 1\}}
\end{equation}
where $(1-L^s)^D$ and $(1-L)^d$ are the differencing operators; $\theta_q(L)$ and $\phi_p(L)$ are lag polynomials having roots all outside the unit circle; $\varepsilon_t$ is white noise with mean 0 and variance $\sigma^2_{\varepsilon}$; $\X_t$ is a set of covariates satisfying the covariates-treatment independence assumption; $\Theta_Q (L^s)$, $\Phi_P(L^s)$ are the lag polynomials of the seasonal part of the model with period $s$ and having roots all outside the unit circle. More importantly, $\tau_t = 0$ $\forall t < t^*$ and $\mathds{1}_{\{\w = 1\}}$ is an indicator function which is 1 if the statistical unit receives the treatment. As highlighted in \citet{Menchetti:Cipollini:Mealli:2021}, this means that $\tau_t$ can be interpreted as the causal effect of the treatment at time point $t$, since it is defined as a contrast of potential outcomes. To see that, indicate with $T(\cdot)$ the transformation of $\Y_t(\w)$ needed to achieve stationarity, i.e. $T(\Y_t(\w)) =(1-L^s)^D (1-L)^d \Y_t(\w) $. Notice that the same transformation is also applied to $\X_t$. Then, we can define
\begin{equation*}
z_t = \frac{\Theta_Q (L^s)\theta_q(L) }{ (1-L^s)^D (1-L)^d \Phi_P(L^s) \phi_p(L)}\varepsilon_t
\end{equation*} 
so that model (\ref{eqn:arima_general}) becomes,
\begin{equation*}
\Y_t(\w) =  z_{t} + \X_{t}' \beta + \tau^Y_{t} \mathds{1}_{\{\w = 1\}} 
\end{equation*}
where 
\begin{equation*}
\tau^Y_t = \frac{\tau_t}{(1-L^s)^D (1-L)^d}
\end{equation*}
is the causal effect on the untrasformed variable, defined as a contrast of potential outcomes, 
i.e., $\tau^Y_t = \Y_t(\w = 1) - \Y_t(\w = 0)$. Similarly, in a multi-intervention setting where $N$ persistent interventions take place at time points $\Lambda \in \{ t_1,\dots,t_N\}$, under non-anticipating treatments and non-anticipating potential outcomes, the outcome at time $t \in \{t_n,\dots,t_{n+1}-1 \}$ is the result of all past interventions,
\begin{equation}
\label{eqn:arima_general_multi}
\Y_{t}(\w) = z_{t} + \X_{t}' \beta + \sum\limits_{j = 1}^{n-1} \tau_{t}^{(j)} + \tau_{t}^{(n)} \mathds{1}_{ \{ \w = 1\} } 
\end{equation}
where the summation comes from the additivity of the effects produced by $n-1$ persistent interventions and $\tau_t^{(n)}$ is the point effect at time $t$ of the $n$-th intervention.  Now, assume we observe $\Y_t$ up to time $t_n-1$ and that $\w = 0$. Then, for a positive integer $k$, the $k$-step ahead forecast of $\Y_t$ in the absence of the $n$-th intervention, given all the information up to time $t_n -1$ is,
\begin{equation}
\label{eqn:1step_forecast}
\hat{\Y}_{t_n-1+k}(0) = E[\Y_{t_n-1+k}(0) | \info{I}{t_n-1}]  = \hat{z}_{t_n-1+k|t_n-1}  + \X_{t_n-1+k}' \beta + \sum\limits_{j = 1}^{n-1} \tau_{t_n-1+k}^{(j)}.
\end{equation}
Notice that $\hat{\Y}_{t_n -1 +k}(0)$ is, by definition, the expectation of the outcome series when the $n$-th intervention does not occur;
thus, it can be considered an estimate of the missing potential outcomes at time $t_n -1 +k$ for a persistent intervention occurring at time $t_n$. Therefore, the point effect of the $n$-th intervention can be estimated by the quantity
\begin{align}
\label{eqn:pred_error}
\nonumber
\hat{\tau}^{(n)}_{t_n-1+k} & = \Y_{t_n -1 + k}(1) - \hat{\Y}_{t_n-1+k}(0) \\ \nonumber
& = z_{t_n -1 + k} + \sum\limits_{j=1}^{n-1} \tau_{t_n -1+k}^{(j)} + \tau_{t_n -1+k}^{(n)} - \hat{z}_{t_n - 1 + k|t_n-1} - \sum\limits_{j = 1}^{n-1} \tau_{t_n -1+k}^{(j)} \\ 
& = z_{t_n -1 + k} - \hat{z}_{t_n - 1 + k|t_n-1} + \tau_{t_n -1+k}^{(n)}. 
\end{align}
Furthermore, the limiting case $k = 1$ leads to the $1$-step ahead forecast $\hat{\Y}_{t_n}$, which is an estimate of the missing potential outcome in the absence of intervention for the contemporaneous effect.

\subsection{Causal effect inference}
\label{subsect:estimators}
Notice that, since Equations (\ref{eqn:1step_forecast})-(\ref{eqn:pred_error}) are defined by conditioning on the information set up to time $t_n-1$, more formally we have $\hat{\Y}_{t_n -1 +k}(0) \equiv \hat{\Y}_{t_n -1 +k}(\w_{1:t_n-1}^{obs}, \bs{0})$ and $\Y_{t_n -1 +k}(1) \equiv \Y_{t_n -1 +k}(\w_{1:t_n-1}^{obs}, \bs{1})$. Based on the C-ARIMA model for the multi-intervention setting, we can now derive estimators for the pointwise and the contemporaneous causal effects. 

\begin{defn}
Denote with $\Lambda = \{t_1,\dots,t_N \}$ the subset of time points at which $N$ persistent interventions take place and indicate with $\w_{1:t_n-1}^{obs}$ the observed treatment path. For a positive integer $k$, let $\Y_{t_n-1+k}(\w_{1:t_n-1}^{obs}, \bs{1})$ be the observed time series and let $\hat{\Y}_{t_n-1+k}(\w_{1:t_n-1}^{obs}, \bs{0})$ be the $k$-step ahead forecast as defined in (\ref{eqn:1step_forecast}). For any $k>0$, an estimator of the point  effect of the $n$-th intervention at time $t_n-1+k$ is,
\begin{equation}
\label{eqn:point_estimator}
\hat{\tau}_{t_n-1+k}^{(n)}(\w_{1:t_n-1}^{obs}, \bs{1}; \bs{0}) = \Y_{t_n-1+k}(\w_{1:t_n-1}^{obs}, \bs{1}) - \hat{\Y}_{t_n-1+k}(\w_{1:t_n-1}^{obs}, \bs{0}) .
\end{equation}
Then, an estimator of the cumulative pointwise effect of the $n$-th intervention up to time $t_n-1+k$ is,
\begin{equation}
\label{eqn:cumpoint_estimator}
\hat{\Delta}_{t_n-1+k}^{(n)}(\w_{1:t_n-1}^{obs},\bs{1}; \bs{0}) = \sum\limits_{h = 1}^{k} \hat{\tau}_{t_n-1+k}^{(n)}(\w_{1:t_n-1}^{obs},\bs{1}; \bs{0})
\end{equation}
and, finally, an estimator of the temporal average pointwise effect is,
\begin{equation}
\label{eqn:avgpoint_estimator}
\hat{\bar{\tau}}_{t_n-1+k}^{(n)}(\w_{1:t_n-1}^{obs},\bs{1}; \bs{0}) = \frac{\hat{\Delta}_{t_n-1+k}^{(n)}(\w_{1:t_n-1}^{obs},\bs{1}; \bs{0})}{k}.
\end{equation}
\end{defn}
Furthermore, by setting $k = 1$ in Equation (\ref{eqn:point_estimator}) we get an estimator of the contemporaneous effect of the $n$-th intervention at time $t_n$,  
\begin{equation*}
\hat{\tau}^{(n)}(\w_{1:t_n-1}^{obs},1; 0) = \Y_{t_n}(1) - \hat{\Y}_{t_n}(0) .
\end{equation*}
Inference on the pointwise effects and, by extension, on the contemporaneous effect can be performed using hypothesis tests based on the above estimators. The following theorem illustrates their distributional
properties.

\begin{theorem}
Let $\{ \Y_t(\w)\}$ follow the regression model with ARIMA errors described by Equation (\ref{eqn:arima_general_multi}) and, for any $k >0$, let $H_0: \tau^{(n)}_{t_n-1+k} = 0$ the null hypothesis that the $n$-th intervention has no effect. Then, for some $b_i$ coefficients with $b_0 = 1$, the estimators of the point, cumulative and temporal average effects under $H_0$ can be expressed as,
\begin{align}
\label{eqn:point_estimator_var_h0}
\hat{\tau}_{t_n-1+k}^{(n)}(\w_{1:t_n-1+k}^{obs};\bs{1},\bs{0})|H_0 & = \sum\limits_{j = 1}^{k} \varepsilon_{t_n-1+j} \sum\limits_{i = 0}^{k-j} b_i \psi_{k-j-i} \\
\label{eqn:cumpoint_estimator_var_h0}
\hat{\Delta}_{t_n-1+k}^{(n)}(\w_{1:t_n-1+k}^{obs}, \bs{1}, \bs{0})|H_0 & = \sum\limits_{h = 1}^k \varepsilon_{t_n-1+h} \sum\limits_{i = 0}^{k-h} b_i \sum\limits_{j = i}^{k-h} \psi_{k-h-j} \\
\label{eqn:avgpoint_estimator_var_h0}
\hat{\bar{\tau}}_{t_n-1+k}^{(n)}(\w_{1:t_n-1+k}^{obs}, \bs{1}, \bs{0})|H_0 & = \frac{1}{k}\sum\limits_{h = 1}^k \varepsilon_{t_n-1+h} \sum\limits_{i = 0}^{k-h} b_i \sum\limits_{j = i}^{k-h} \psi_{k-h-j}, 
\end{align}
where the $\psi$'s are the coefficients of a moving average of order $k-1$ whose values are function of the ARMA parameters in Equation (\ref{eqn:arima_general_multi}). In case the error term $\varepsilon_t$ is assumed to be Normally distributed, Equations (\ref{eqn:point_estimator_var_h0})--(\ref{eqn:avgpoint_estimator_var_h0}) become,

\begin{align}
\label{eqn:point_estimator_var}
\hat{\tau}_{t_n-1+k}^{(n)}(\w_{1:t_n-1}^{obs},\bs{1};\bs{0})|H_0 
& \sim 
N \left[ 0, \sigma^2_{\varepsilon_{n-1}}\sum\limits_{j = 1}^{k} \left( \sum\limits_{i = 0}^{k-j} b_i \psi_{k-j-i} \right)^2 \right] \\
\label{eqn:cumpoint_estimator_var}
\hat{\Delta}_{t_n-1+k}^{(n)}(\w_{1:t_n-1+k}^{obs}, \bs{1}, \bs{0})|H_0 
& \sim N \left[ 0,  \sigma^2_{\varepsilon_{n-1}}\sum\limits_{h=1}^k  \left(\sum\limits_{i = 0}^{k-h} b_i \sum_{j=i}^{k-h} \psi_{k-h-j} \right)^2 \right] \\
\label{eqn:avgpoint_estimator_var}
\hat{\bar{\tau}}_{t_n-1+k}^{(n)}(\w_{1:t_n-1+k}^{obs}, \bs{1}, \bs{0})|H_0 
& \sim N \left[ 0, \frac{\sigma^2_{\varepsilon_{n-1}}}{k^2}\sum\limits_{h=1}^k  \left(\sum\limits_{i = 0}^{k-h} b_i \sum_{j=i}^{k-h} \psi_{k-h-j} \right)^2 \right],
\end{align}

where $\sigma_{\varepsilon_{n-1}}$ is the variance of the error term of the model estimated on the observations up to time $t_n-1$.

Proof: the derivation is analogous to Theorem 2 in \citet{Menchetti:Cipollini:Mealli:2021}, where we can replace the single intervention date $t^*$ with $t_n-1$, the day before the $n$-th intervention.

\end{theorem}


Thus, the estimation of the pointwise effect is performed in two steps.  First, we estimate a C-ARIMA model up to the day preceding the persistent intervention, i.e., the launch of the two futures by CBOE and CME. Then, we use the covariates and the estimated coefficients to forecast the outcome time series (i.e., Garman-Klass volatility proxy and trading volume) up to a pre-specified time horizon; for example, we may be interested in estimating the effect on volatility after few days or few months from the launch of Bitcoin futures. The difference between the observed and the predicted outcome is the estimated pointwise effect. Similarly, we can estimate the contemporaneous effects of each announcement by conditioning to the information set up to the day before the announcement and forecasting the Garman-Klass proxy and the trading volume one-step ahead. If the announcements or the launch of futures had an impact on Bitcoin volatility and volume, we would find a significant deviation from the forecasted outcomes.

%
%
%

\section{Empirical application}
\label{sect:data_method}
\subsection{Data}
Economic data for this analysis have been gathered from Bloomberg, while Bitcoin daily prices have been collected from CryptoCompare. Since Bitcoin is a (crypto)currency, its price is recorded in terms of another currency, USD in our case, meaning that the Bitcoin price is the BTC-USD exchange rate (number of USD needed to buy 1 BTC). The other Bitcoin related data (i.e. the hash rate and total number of Bitcoins in circulation) have been gathered from Blockchain.com. 

Since Bitcoins are traded on multiple exchanges, the quotation of the BTC-USD rate is not unique, i.e. different exchanges trade Bitcoins at different prices. As explained in \citet{Cermak:2017}, the price varies across exchanges mainly because of different fee policies and cashout methods but those divergent standards and the slow verification process make arbitrage opportunities difficult to exploit\footnote{For example, on November 20, 2019 at 4.50 p.m the price for 1 BTC was \$8,104.39 on Coinbase and \$8,146.90 on Bitfinex, with a difference of \$42.51 (Source: CryptoCompare.com).}. Aside from the obvious economic implications, this means that the source we take our data from matters: exchange-based data providers present their own quotes, whereas external data providers (e.g. Bloomberg, CryptoCompare) compute their own index, usually a weighted average of all prices across major exchanges.

The goal of our analysis is estimating the effect of futures trading on Bitcoin volatility and volumes, and, since the upcoming futures were announced by several press releases, we also investigate possible announcement effects. We focus on the first two futures introduced by CBOE and CME, hence, the analysis period spans from May 2014 to January 2018. We start from May 2014 to avoid the market turbulence following the failure of Mt. Gox; ending the analysis period in January 2018 allows us to follow the Bitcoin network for three weeks after the launch of the CME future. 

\begin{figure}[h!]
\centering
\caption{Garman-Klass volatility proxy in log scale. Panel (A) displays the time series evolution during the analysis period; Panel (B) shows the evolution of the time series starting from June 2017 and highlights the announcement dates and the launch of the two futures (see Table \ref{tab:event_history} for the details and the exact dates); Panel (c) shows the Normal QQ Plot, autocorrelation and partial autocorrelation functions.}
\label{fig:series}
\begin{tabular}{m{1cm}m{15cm}}
A) & \includegraphics[scale=0.35]{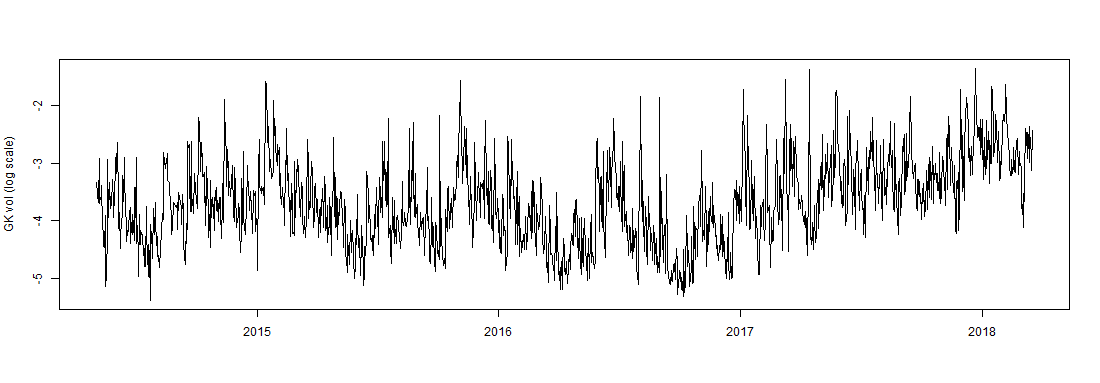} \\
B) & \includegraphics[scale=0.35]{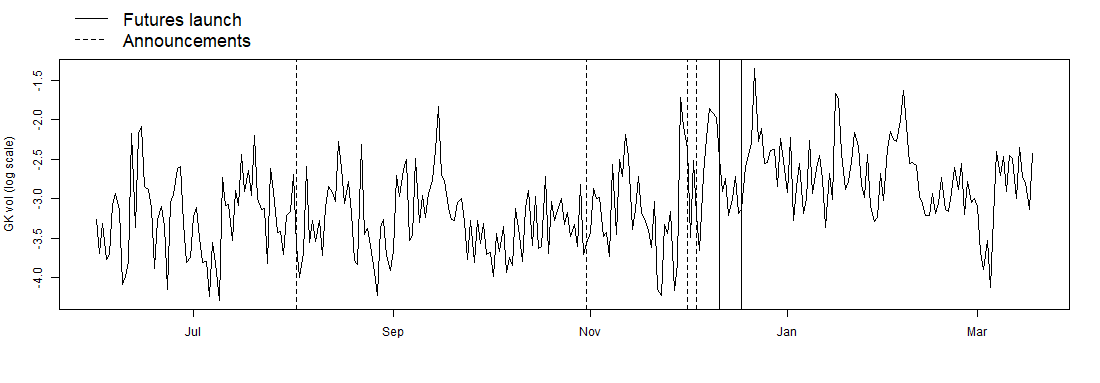} \\
C) & \includegraphics[scale=0.35]{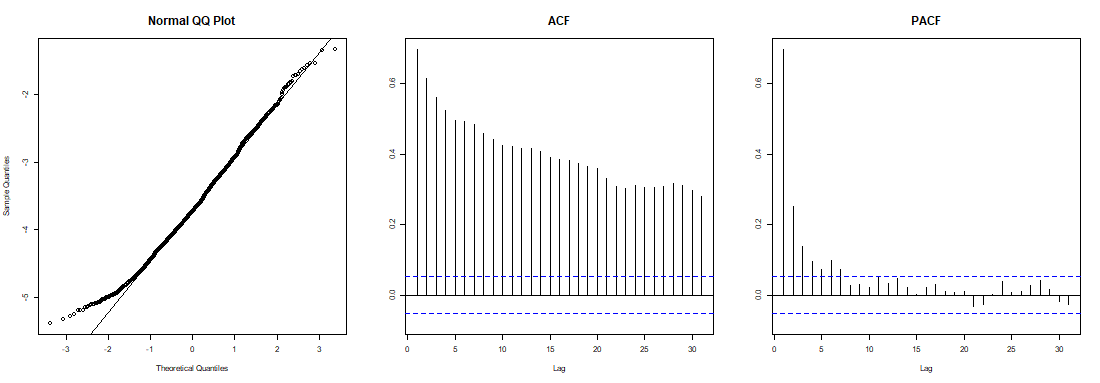} \\
\end{tabular}
\end{figure}

The BTC-USD daily volatility is proxied by the unbiased Garman-Klass estimator  \citep{Garman:Klass:1980,Molnar:2012}, computed as
\begin{align*}
\hat{\sigma}_{GK}^2 & = 0.5 (\ln(H)-\ln(L))^2 - (2\ln2 -1)(\ln(C) - \ln(O))^2 \\
\hat{\sigma}_{GK} & = \sqrt{\hat{\sigma}_{GK}^2} \cdot 1.034 
\end{align*}
where $H,L,C$ and $O$ indicate, respectively, the high, low, close and open BTC-USD rate for the day. In our application, the dependent variable is the natural logarithm of the Garman-Klass proxy.  Figure \ref{fig:series} shows its evolution during the analysis period, the Normal QQ plot and the (partial) autocorrelation function.  

\begin{figure}[h!]
\centering
\caption{Bitcoin daily transaction volumes in log scale. Panel (A) displays the time series evolution during the analysis period; Panel (B) shows the evolution of the time series starting from February 2015 and highlights the announcement dates and the launch of the two futures (see Table \ref{tab:event_history} for the details and the exact dates); Panel (c) shows the Normal QQ Plot, autocorrelation and partial autocorrelation functions.}
\label{fig:volumes}
\begin{tabular}{m{1cm}m{15cm}}
A) & \includegraphics[scale=0.35]{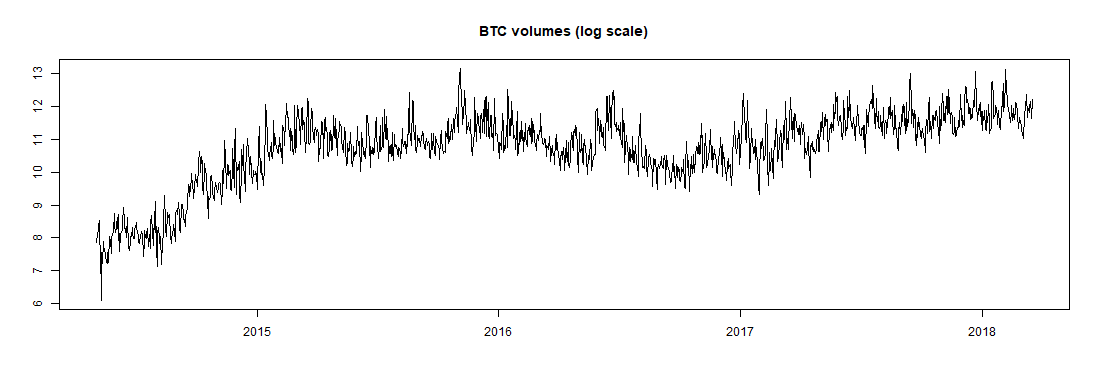} \\
B) & \includegraphics[scale=0.35]{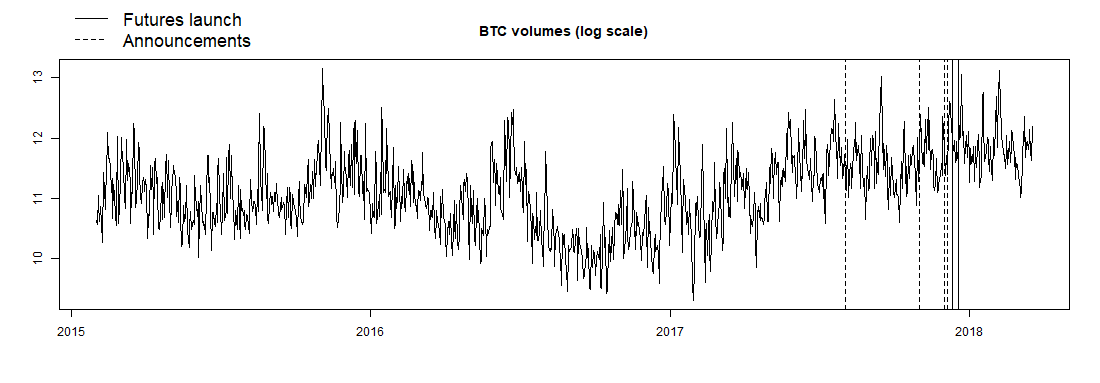} \\
C) & \includegraphics[scale=0.35]{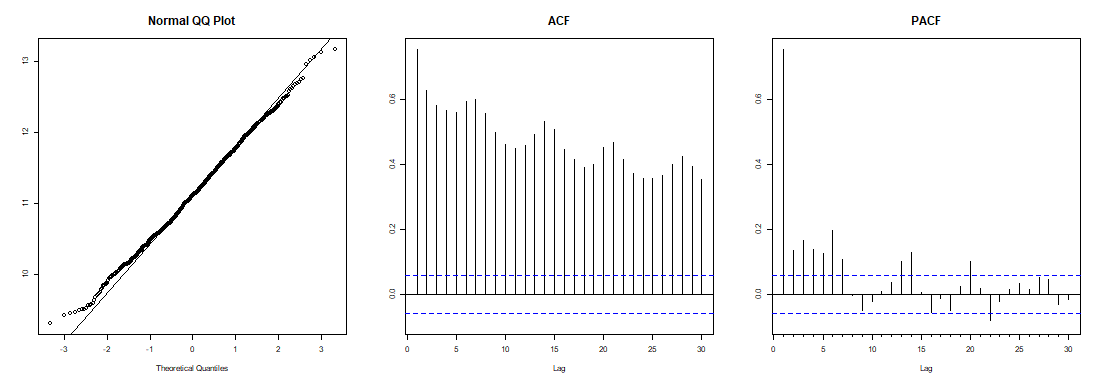} \\
\end{tabular}
\end{figure}

In order to shed light on the driving forces underneath Bitcoin volatility, we also investigate the impact of Bitcoin futures on the daily transaction volumes. Indeed, the possible volatility surge (decline) following the launch of the two contracts might be driven by increased (decreased) volumes. From Figure \ref{fig:volumes}, showing the evolution of daily volumes throughout the analysis period, we can notice that Bitcoin transactions experienced a sharp increase during 2014. However, as shown in Panel (B), by focusing on a restricted time period, the time series becomes more stable. 

\clearpage
\subsection{Methodology}
\label{subsect:method}
As described in Section \ref{subsect:estimators}, to estimate the effect on Bitcoin volatility and volumes of the two futures and each related announcement, we need to fit a C-ARIMA model up to the day before each intervention, which, in our case, leads to the estimation of  six different models: based on the first four, we make $1$-step ahead predictions to compute the contemporaneous effect of each announcement, whereas, based on the last two models, we compute the effect of futures at different time horizons. In particular, since the CBOE future was launched $1$ week before the CME future, for the former we estimate the temporal average pointwise effect at $1$-week horizon, whereas for the CME futures we can also consider the $2$-week and the $3$-week horizons. 

Notice that since both dependent variables are in log scale, the difference between the observed and the predicted outcomes is equal to the log of their ratio. This means that we are assuming a multiplicative effect and, as a result, among the estimators defined by Equations (\ref{eqn:point_estimator})--(\ref{eqn:avgpoint_estimator}) we can focus on the temporal average pointwise effect. Indeed, it has a financial interpretation after re-exponentiating, being the geometric average of the point effects.

Volatility is typically stationary and non-seasonal, hence, our six independent models are all built from an ARMA$(p,q)$. The models also include covariates, so to improve the forecast of Bitcoin volatility in the absence of intervention. We selected the set of covariates based on a survey of relevant literature (see Section \ref{sect:background}). In particular, we included the daily log returns and the Garman-Klass proxies of the EUR-USD exchange rate, the MSCI Emerging Market Index, the Shanghai Stock Exchange Composite Index and the MSCI World Index. Then, among the economic factors we selected the Federal Reserve money supply M1 aggregate, the US monthly inflation rate, the Federal funds Target Rate and the US GDP. Technology factors are represented by the hash rate (estimated number of tera hashes per second the Bitcoin network is performing) and Bitcoin supply (total number of Bitcoins in circulation). All covariates are in log scale and have been made stationary in case they are not\footnote{First differences have been taken for \texttt{m1}, \texttt{midrate}, \texttt{gdp} and \texttt{hash}, whereas \texttt{btcsupply} has been differenced twice.}. Finally, since the second Bitcoin halving took place on July 9, 2016, we also included a dummy variable taking value $1$ after the halving. Table \ref{tab:corr} summarizes the correlation between the covariates and the Garman-Klass volatility. Overall, the Garman-Klass proxy seems to have a small correlation with the volatility of the EUR-USD exchange rate. All the considered covariates are reasonably unaffected by Bitcoin futures; thus, we can use their values post-intervention to improve the prediction accuracy of the counterfactual outcomes absent the futures.

The selection of the six independent models was based on the Bayesian Information Criterion (BIC). Since the announcement dates are close to each other and to the actual launch of the futures, the characteristics of the data are in all cases well described by the same C-ARMA$(2,1)$ specification, i.e.,\footnote{Notice from Figure \ref{fig:series} that the acf of the Bitcoin Garman-Klass volatility indicator shows a very slow decay, a behavior which is sometimes labeled as \textit{long-memory}.
To capture this pattern, specific models have been developed, like the HAR (Heterogeneus Autoregressive) and its log-counterpart, the log-HAR \citep{Corsi:2009}. On the other hand, \citet[][Sect.~5.3]{Cipollini:Gallo:Palandri:2020} demonstrate that some (2,1)-specifications of different ARMA-like models replicate the ability to approximate the long memory pattern observed in the autocorrelation of realized variances, a feature which has made the HAR model popular.
Such an ability is due to the high estimated \textit{persistence} and the presence of a second order parameter, usually significantly negative, in the AR part.}
\begin{equation}
\label{eqn:volatility}
\Y_t(\w) = \frac{\theta_1(L)}{\phi_2(L)}\varepsilon_t + \X_t' \beta + \sum_{j = 1}^{n-1} \tau_t^{(j)} + \tau_t^{(n)}\mathds{1}_{\{\w=1 \}} .
\end{equation}
Unlike the Garman-Klass proxy, log-volumes are not stationary (the KPSS test rejects the null hypothesis of stationarity at the $1\%$ level) and show a weekly seasonal pattern; thus, denoting with $\V_t(\w)$ the potential outcome time series of log-volumes at time $t$, the model selected with the BIC criterion is a C-ARIMA$(1,1,1)(0,0,2)_7$, which is written as,
\begin{equation}
\label{eqn:volumes}
\V_t(\w) = \frac{\Theta_2(L^7) \theta_1(L)}{(1-L)\phi_1(L)}\varepsilon_{t} + \X_t' \beta + \sum_{j = 1}^{n-1} \tau_t^{(j)} + \tau_t^{(n)}\mathds{1}_{\{\w=1 \}}.
\end{equation}
In case that trading volumes turn out to be affected by the two futures and if the volume-volatility relationship holds also in the Bitcoin market, the effect on Bitcoin volatility estimated by (\ref{eqn:volatility}) would then be due, at least in part, by increased (decreased) transactions.
To find the impact on Bitcoin volatility attributable solely to the two futures, we should be able to predict the counterfactual outcome that we would have had absent the futures and their impact on volumes. We can do that by using the following new model for the Garman-Klass volatility proxy:
\begin{equation}
\label{eqn:mixed_model}
\Y_t(\w) = \frac{\theta_q(L)}{\phi_p(L)}\varepsilon_{t} + \X_t' \beta + (1-L) \sum_{i = 1}^3 \alpha_i \tilde{\V}_{t-i} + \sum_{j = 1}^{n-1} \tau_t^{(j)} + \tau_t^{(n)}\mathds{1}_{\{\w=1 \}}
\end{equation} 
where $\tilde{\V}_{t} = (\V_1, \dots, \V_{t_n-1},\hat{\V}_{t_n},\dots,\hat{\V}_T)$ is a vector formed by the observed volumes up to the day before the $n$-th intervention and then by the counterfactual volumes forecasted by model (\ref{eqn:volumes}), denoted with $\hat{\V}_t$; thus, $\tilde{\V}_{t-1}$, $\tilde{\V}_{t-2}$ and $\tilde{\V}_{t-3}$ are the lagged values of $\tilde{\V}_{t}$.\footnote{Notice that including $\tilde{\V}_t$ in the set of covariates does not violate the covariates-treatment independence assumption. Indeed, by using the predicted counterfactual volumes $\hat{\V}_t$ in the post intervention period, $\tilde{\V}_t$ is truly unaffected by the interventions. } The idea is that, in the pre-intervention period, significant coefficients in the lagged volumes would confirm the presence of a volume-volatility relationship in the Bitcoin market, whereas, in the post-intervention period, the lagged predicted volumes would improve the estimation of the counterfactual volatility absent the futures and their effect on volumes. The next Section reports the results of the empirical analysis on the Garman-Klass volatility proxy and on Bitcoin daily volumes.

\section{Results}
\label{sect:results}
\begin{figure}[h!]
\centering
\caption{Observed and forecasted \textbf{Garman-Klass volatility proxy} (in log scale) at $1$-week, $2$-weeks and $3$-weeks from the launch of the CME future, indicated by the dashed bar. The right charts show the resulting effects (computed as the difference between the observed and the forecasted series) with their $95 \%$ confidence bounds.}
\label{fig:pointwise_cme}
\begin{tabular}{cc}
\includegraphics[scale=0.40]{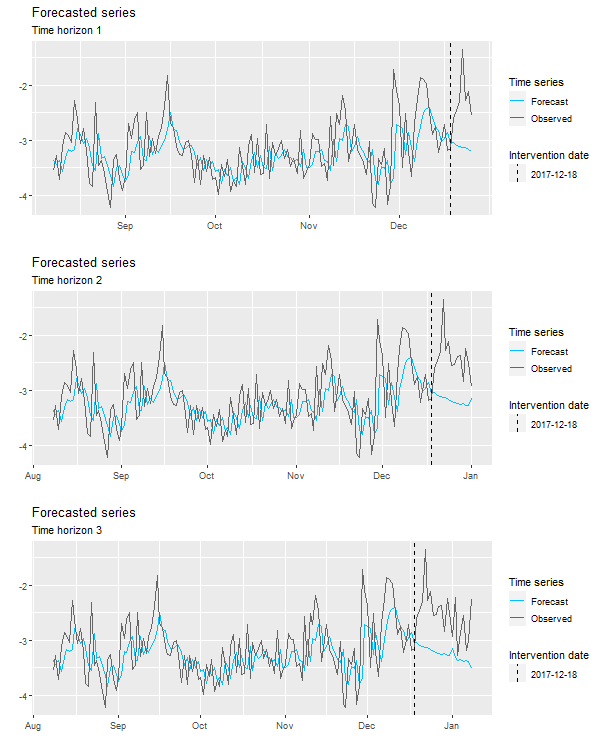} & \includegraphics[scale=0.40]{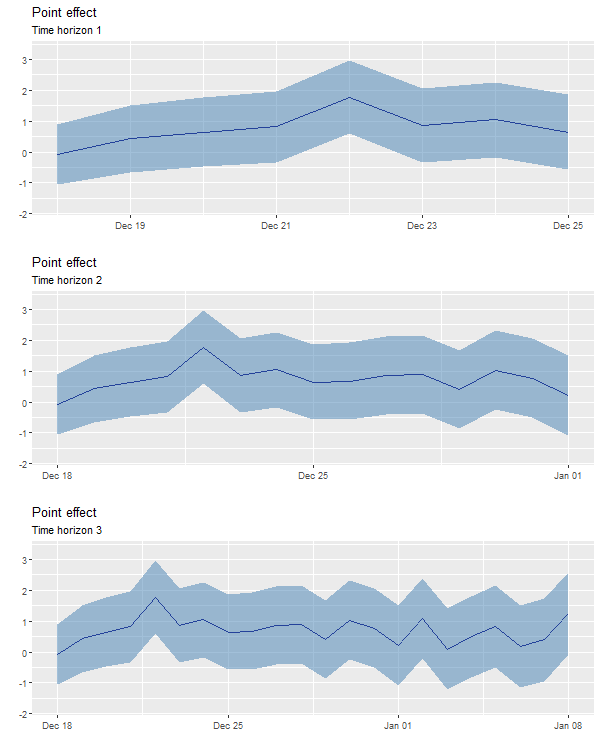} \\
\end{tabular}
\end{figure}

The parameter estimates of the six C-ARMA$(2,1)$ models, the estimated contemporaneous effect of the announcements and the estimated temporal average pointwise effects of the two futures on the Garman-Klass volatility proxy are reported in Table \ref{tab:estimates}. Since the residuals diagnostics do not seem to support the Normality assumption (see Figure \ref{fig:residuals_GK}), we report empirical critical values computed from Equation (\ref{eqn:avgpoint_estimator_var_h0}) by bootstrapping the error terms from the model residuals.\footnote{The number of bootstrap samples is $10,000$. See Table \ref{tab:sd_norm} for the estimated standard errors of the temporal average pointwise effects computed from Equation (\ref{eqn:avgpoint_estimator_var}).} 
We find no evidence of significant contemporaneous effects of the announcements, meaning that the news of upcoming futures was not sufficient to spark investors' interests in the (crypto)currency. Instead, we observe an interesting result for the actual introduction of the two futures: there is no evidence of a significant effect related to CBOE futures, whereas the effects related to CME futures are significant at all time horizons. For example, we find that, $1$-week after its launch, the CME future increased volatility by, on average, $+116\%$. 
The analysis on Bitcoin volumes might provide an explanation of the different results found for the CBOE and CME instruments.
Figure \ref{fig:pointwise_cme} shows the forecasted series and the pointwise causal effects. Having found no evidence of association between the covariates and the Garman-Klass proxy (lagged values of the volatility seem sufficient in describing the dynamics of current Bitcoin volatility), as a robustness check to the discussed results we also estimated six alternative models with no regressors. The estimated causal effects based on the alternative models are in line with the results reported in this section (see Table \ref{tab:estimates_2} for the details). 

\begin{figure}[h!]
\centering
\caption{Observed and forecasted \textbf{Bitcoin daily volumes} (in log scale) at $1$-week, $2$-weeks and $3$-weeks from the launch of the CME future, indicated by the dashed bar. The right charts show the resulting effects (computed as the difference between the observed and the forecasted series) with their $95 \%$ confidence bounds.}
\label{fig:pointwise_cme_vol}
\begin{tabular}{cc}
\includegraphics[scale=0.4]{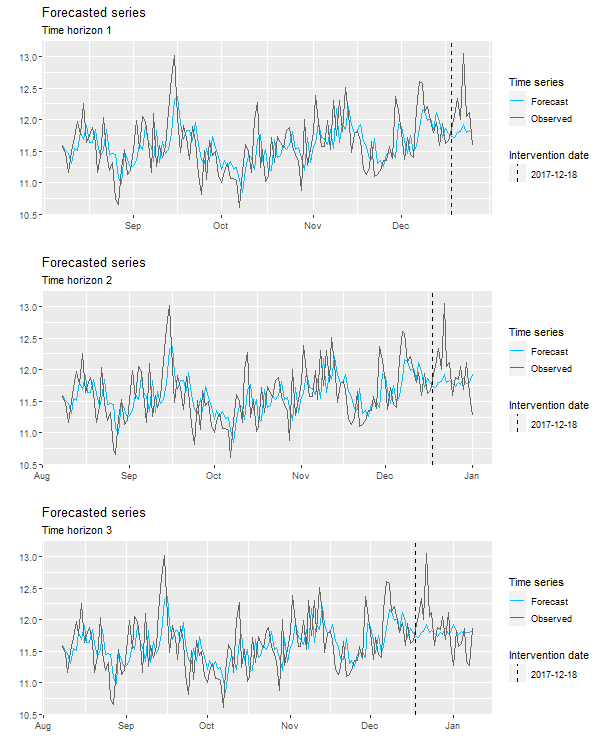} & \includegraphics[scale=0.4]{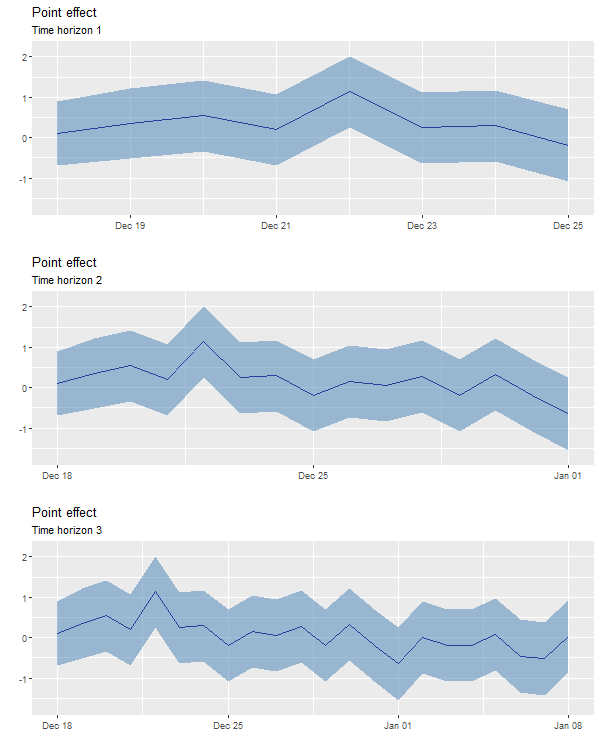} \\
\end{tabular}
\end{figure}

Table \ref{tab:estimates_volumes_log} shows the results of the analysis performed on Bitcoin daily transaction volumes (in log scale). Again, we report empirical critical values computed by bootstrapping the errors.\footnote{The number of bootstrap samples is $10,000$. See Table \ref{tab:sd_norm} for the estimated standard errors of the temporal average pointwise effects computed from Equation (\ref{eqn:avgpoint_estimator_var}).} This time we found significant effects for both the CBOE and the CME futures at all time horizons, whereas there is no evidence of a contemporaneous causal effect of the announcements. Trading volumes show a positive association with Bitcoin supply, meaning that the growing number of coins also translates into more transactions. In addition, Bitcoin volumes seem to be negatively related to the MSCI World Index, suggesting that the cryptocurrency might act as a safe heaven during turbulence in traditional markets. Figure \ref{fig:pointwise_cme_vol} provides a graphical representation of these results. As a robustness check, we also repeated the analyses on a shorter time interval starting from February 2015 when Bitcoin volumes, after a sharp increase during 2014, finally reached a more stable path. The results are in line with those reported in this section and are shown in Table \ref{tab:estimates_volumes_2}. Interestingly, the effects of the two futures have opposite sign: transaction volumes decreased of approximately $-18\%$ due to the launch of the CBOE contract, which might indicate that for a short period of time investors privileged the derivative over of the underlying. Lower transaction volumes might also explain the absence of effect on volatility. Conversely, the CME contract increased volumes by $+39\%$ in the first week after its launch, suggesting that, benefiting from the increased transparency of the market, investors' interest toward Bitcoin rose (especially in the short-run) boosting Bitcoin trading and, in turn, its volatility. 

To check whether the above interpretation is correct and to find the impact on volatility due solely to the futures, we estimated one last model for the Garman-Klass volatility proxy where we included a few lagged values of transaction volumes in the set of predictors, as in Equation (\ref{eqn:mixed_model}): if the effect on volatility was, at least in part, triggered by volumes, we would expect to find significant coefficients for the lagged log-volumes (since volumes exhibit non-stationarity, we considered their first difference). Then, as described in Section \ref{subsect:method}, to improve the prediction of the counterfactual volatility absent the futures and their impact on volumes, in the post-intervention period we used $\hat{\V_t}$. Table \ref{tab:mediation} reports the results: as expected, the coefficient of lagged volumes is positive and significant, suggesting that when transactions increase Bitcoin volatility rises as well. Interestingly, past volumes seem to exert a strong influence on volatility even at $t-3$.\footnote{We also tested the presence of possible feedback effects stemming from volatility to volumes. To do that, in Equation (\ref{eqn:mixed_model}) we switched lagged log-volumes with lagged log-volatilities and $\Y_t$ with $\V_t$. As shown in Table \ref{tab:mediation_volatility}, we did not find evidence of a feedback effect.} Notice that by considering the volumes that we would have observed in the absence of intervention, the estimated causal effect on Bitcoin volatility reported in Table \ref{tab:mediation} is the actual impact due solely to the two futures, which is then higher than the effect reported in Table \ref{tab:estimates}. For example, we obtain that, considering the lower volumes that we would have had in the absence of the future, the CME contract is actually responsible of a $+139\%$ increase in volatility in first week of its launch
(instead of the $+116\%$ estimated by the first model).\footnote{This is because, by considering lower transactions, the counterfactual volatility predicted by model (\ref{eqn:mixed_model}) is below the one predicted by model (\ref{eqn:volatility}).}  Again, the effect of the CBOE contract was not significant, whereas the causal effect of the CME contract is significant at all time horizons, with a small reduction after $3$ weeks.

\begin{table}[h!]
\centering
\caption{Estimates of the six independent C-ARMA models fitted on the historical daily values of the \textbf{Garman-Klass volatility proxy} (in log scale) starting from May 3, 2014 and up to the day before each intervention (standard errors within parentheses). See Table \ref{tab:event_history} for the exact dates of the announcements and futures launches. In this table, $\hat{\tau}^{(n)}$ indicates the estimated contemporaneous effect of each announcement; $\hat{\tau}^{(CBOE)}$ is the temporal average pointwise effect of the CBOE futures; and, $\hat{\tau}_t^{(CME)}$ is the temporal average pointwise effect of the CME futures at $1$-week, $2$-weeks and $3$-weeks horizons, respectively $t = 7$, $t = 14$ and $t = 21$. The multiplicative effects can be recovered by re-exponentiating the estimated effects. The empirical critical values for the causal effects are computed from Equation (\ref{eqn:avgpoint_estimator_var_h0}) by bootstrapping the errors from the residuals of each model.}
\label{tab:estimates}
\scalebox{0.75}{
\input{Tables/BTC_gk_est}
}
\end{table}

\begin{table}[ht]
\centering
\caption{Estimates of six C-ARIMA models fitted on the historical \textbf{Bitcoin daily volumes} (in log scale) starting from May 3, 2014 and up to the day before each intervention (standard errors within parentheses). See Table \ref{tab:event_history} for the exact dates of the announcements and futures launches. In this table, $\hat{\tau}^{(n)}$ indicates the estimated contemporaneous effect of each announcement; $\hat{\tau}^{(CBOE)}$ is the temporal average pointwise effect of the CBOE futures; and, $\hat{\tau}_t^{(CME)}$ is the temporal average pointwise effect of the CME futures at $1$-week, $2$-weeks and $3$-weeks horizons, respectively $t = 7$, $t = 14$ and $t = 21$). The multiplicative effects can be recovered by re-exponentiating the estimated effects. The empirical critical values for the causal effects are computed from Equation (\ref{eqn:avgpoint_estimator_var_h0}) by bootstrapping the errors from the residuals of each model.}
\label{tab:estimates_volumes_log}
\scalebox{0.75}{
\input{Tables/BTC_volumes_est}
}
\end{table}

\begin{table}
\centering
\caption{Estimates of the C-ARMA models fitted on the historical daily values of the \textbf{Garman-Klass volatility proxy} (in log scale) with lagged volumes among regressors (standard errors within parentheses). The models are estimated starting from May 3, 2014 and up to the day before each intervention. See Table \ref{tab:event_history} for the exact dates of the futures launches. In this table, $\hat{\tau}^{(CBOE)}$ is the temporal average pointwise effect of the CBOE futures and $\hat{\tau}_t^{(CME)}$ is the temporal average pointwise effect of the CME futures at $1$-week, $2$-weeks and $3$-weeks horizons (indicated with $t = 7$, $t = 14$ and $t = 21$). The multiplicative effects can be recovered by re-exponentiating the estimated effects. The empirical critical values for the causal effects are computed from Equation (\ref{eqn:avgpoint_estimator_var_h0}) by bootstrapping the errors from the residuals of each model.}
\label{tab:mediation}
\scalebox{0.75}{
\input{Tables/BTC_mediation}

}
\end{table}

\newgeometry{bottom=1.5cm,top=1.5cm,left=1cm,right=1cm}

\restoregeometry

\newgeometry{bottom=2.5cm,top=2.3cm,left=2cm,right=2cm}
\section{Conclusions}
\label{sect:conclusions}
In December 2017, two leading derivative exchanges, CBOE and CME, introduced the first two regulated futures having Bitcoin as underlying asset. Contributing to the nascent stream of literature seeking to assess the effect generated by the new instruments, the goal of this paper was estimating their causal impact on Bitcoin volatility; we also investigated the presence of announcement effects as well as the impact on Bitcoin trading volumes.

To estimate the causal effect generated by the launch of Bitcoin futures, 
we employed a novel methodology, C-ARIMA, based on the Rubin's potential outcomes framework. After a detailed discussion of the assumptions enabling the definition of the effect and its attribution to the a specific treatment, this approach infers the causal effect of an intervention by direct comparison between the observed outcome and a predicted counterfactual. In this paper, we extended the C-ARIMA approach to a multi-intervention setting by formalizing an additional assumption and introducing two causal estimands capturing the effect of the multiple announcements and of the futures introductions. Then, we defined estimators of these effects under the C-ARIMA model and derived hypothesis tests. 

The results indicate that the CME contract produced an increase of both Bitcoin volatility and trading volume. We also found evidence that lagged volumes are positively associated to Bitcoin volatility, suggesting that the effect of the introduction of CME futures on volatility might be explained, at least partially, by the increased volumes following the same intervention. After controlling for the lagged predicted volumes in the absence of intervention, we were able to find the effect on volatility actually attributable to the future contract. This effect was positive and significant at all time horizons. Conversely, the CBOE futures had no impact on volatility and produced a small negative effect on volumes. Finally, we did not find evidence of announcement effects. 

Overall, our results support the general finding in the literature of a positive association between volumes and volatility. Furthermore, they show that future trading (although in a regulated environment) contributed to increase Bitcoin volatility instead of diminishing it. Whether this is the result of improved transparency in the market (promoting transactions and hence producing more volatile returns) or increased speculation, this important evidence is backed up by a reliable causal approach: regulators should then be aware of these mechanisms as banks and financial institutions get increasingly involved in the crypto world. 

\restoregeometry

\newpage

\bibliography{C:/Users/fiamm/Dropbox/references2}

\newpage

\begin{appendices}

\section{}
\label{appA}
\numberwithin{figure}{section}
\numberwithin{table}{section}
\setcounter{figure}{0}
\setcounter{table}{0}

\subsection{Additional tables}
\label{appA_tables}

\begin{table}[ht]
\centering
\caption{Estimates of six alternative \textbf{C-ARIMA models without regressors} fitted on the historical daily values of the Garman-Klass volatility proxy (in log scale) starting from May 3, 2014 and up to the day before each intervention (standard errors within parentheses). See Table \ref{tab:event_history} for the exact dates of the announcements and futures launches. In this table, $\hat{\tau}^{(n)}$ indicates the estimated contemporaneous effect of each announcement; $\hat{\tau}^{(CBOE)}$ is the temporal average pointwise effect of the CBOE futures; and, $\hat{\tau}_t^{(CME)}$ is the temporal average pointwise effect of the CME futures at $1$-week, $2$-weeks and $3$-weeks horizons, respectively $t = 7$, $t = 14$ and $t = 21$. The multiplicative effects can be recovered by re-exponentiating the estimated effects. The empirical critical values for the causal effects are computed from Equation (\ref{eqn:avgpoint_estimator_var_h0}) by bootstrapping the errors from the residuals of each model.}
\label{tab:estimates_2}
\scalebox{0.75}{
\input{Tables/BTC_gk_est_app}
}
\end{table}

\clearpage

\begin{table}[ht]
\centering
\caption{Estimates of six C-ARIMA models fitted on the historical \textbf{Bitcoin daily volumes} (in log scale) starting from February 1, 2015 and up to the day before each intervention (standard errors within parentheses). See Table \ref{tab:event_history} for the exact dates of the announcements and futures launches. In this table, $\hat{\tau}^{(n)}$ indicates the estimated contemporaneous effect of each announcement; $\hat{\tau}^{(CBOE)}$ is the temporal average pointwise effect of the CBOE futures; and, $\hat{\tau}_t^{(CME)}$ is the temporal average pointwise effect of the CME futures at $1$-week, $2$-weeks and $3$-weeks horizons, respectively $t = 7$, $t = 14$ and $t = 21$). The multiplicative effects can be recovered by re-exponentiating the estimated effects. The empirical critical values for the causal effects are computed from Equation (\ref{eqn:avgpoint_estimator_var_h0}) by bootstrapping the errors from the residuals of each model.}
\label{tab:estimates_volumes_2}
\scalebox{0.75}{
\input{Tables/BTC_volumes_est_app}
}
\end{table}

\clearpage

\begin{table}
\centering
\caption{Estimates of the C-ARIMA models fitted on the historical \textbf{Bitcoin daily volumes} (in log scale) with lagged volatilities among regressors (standard errors within parentheses). The models are estimated starting from May 3, 2014 and up the day before each intervention. See Table \ref{tab:event_history} for the exact dates of the futures launches.}
\label{tab:mediation_volatility}
\scalebox{0.75}{
\input{Tables/BTC_mediation_volatility}
}
\end{table}

\begin{table}
\centering
\caption{Causal effect estimates of the announcements and futures launches on Bitcoin volatility and volumes. The standard errors in parentheses are estimated from Equation (\ref{eqn:avgpoint_estimator_var}) under the Normality assumption. Panel (1) reports the estimated effects on volatility based on Equation (\ref{eqn:volatility}); Panel (2) reports the estimated effects on volumes based on Equation (\ref{eqn:volumes}); Panel (3) reports the estimated effects on volatility generated by the launch of the two futures based on Equation (\ref{eqn:mixed_model}). In this table, $\hat{\tau}^{(n)}$ with $n = 1,2,3,4$ indicates the estimated contemporaneous effect of each announcement; $\hat{\tau}^{(CBOE)}$ is the temporal average pointwise effect of the CBOE futures; and, $\hat{\tau}_t^{(CME)}$ is the temporal average pointwise effect of the CME futures at $1$-week, $2$-weeks and $3$-weeks horizons, respectively $t = 7$, $t = 14$ and $t = 21$. The multiplicative effects can be recovered by re-exponentiating the estimated effects.}
\label{tab:sd_norm} 
\input{Tables/BTC_norm}
\end{table}

\begin{sidewaystable}[ht]
\centering
\caption{Pearson's linear correlation coefficient (upper triangular) and Spearman's rho (lower triangular) between Garman-Klass volatility (GK) and the covariates included in the analysis, computed in the period before the first intervention.}
\label{tab:corr}
\scalebox{0.8}{
\begin{tabular}{lrrrrrrrrrrrrrrrr} \toprule
 & GK & eurusd\_vol & mxwo\_vol & mxef\_vol & shc\_vol & eurusd & mxwo & mxef & shc & m1 & inflation & gdp & midrate & hash & tot.btc & halv \\ \midrule
GK &  & 0.06 & -0.04 & -0.03 & 0.02 & 0.02 & -0.01 & 0.01 & 0.01 & -0.00 & 0.00 & 0.01 & 0.02 & 0.01 & -0.01 & -0.00 \\ 
  eurusd\_vol & 0.04 &  & 0.47 & 0.43 & 0.35 & 0.03 & -0.03 & -0.05 & -0.02 & -0.03 & 0.01 & -0.00 & 0.05 & -0.04 & -0.07 & -0.16 \\ 
  mxwo\_vol & -0.06 & 0.44 &  & 0.67 & 0.27 & 0.02 & -0.10 & -0.11 & -0.07 & -0.03 & -0.04 & -0.00 & 0.04 & -0.04 & -0.06 & -0.28 \\ 
  mxef\_vol & -0.05 & 0.41 & 0.63 &  & 0.37 & 0.00 & -0.07 & -0.09 & -0.08 & -0.00 & -0.02 & -0.05 & 0.06 & -0.02 & -0.01 & -0.22 \\ 
  shc\_vol & 0.02 & 0.37 & 0.30 & 0.39 &  & -0.03 & -0.05 & -0.09 & -0.12 & 0.02 & -0.01 & -0.04 & -0.03 & -0.01 & 0.04 & -0.55 \\ 
  eurusd & 0.01 & 0.02 & 0.02 & 0.00 & -0.01 &  & -0.03 & -0.04 & -0.10 & 0.00 & 0.00 & -0.05 & -0.01 & -0.02 & 0.01 & 0.05 \\ 
  mxwo & 0.04 & -0.01 & -0.10 & -0.04 & -0.02 & 0.01 &  & 0.65 & 0.19 & -0.04 & 0.01 & 0.01 & 0.05 & -0.01 & -0.02 & 0.04 \\ 
  mxef & 0.02 & -0.02 & -0.08 & -0.06 & -0.09 & -0.04 & 0.52 &  & 0.36 & -0.02 & -0.01 & 0.03 & 0.04 & 0.03 & -0.03 & 0.06 \\ 
  shc & 0.00 & 0.01 & -0.03 & -0.01 & -0.04 & -0.06 & 0.12 & 0.26 &  & -0.02 & 0.04 & 0.05 & -0.00 & -0.02 & 0.01 & -0.01 \\ 
  m1 & -0.01 & -0.04 & -0.03 & 0.01 & 0.03 & 0.05 & -0.05 & -0.01 & 0.01 &  & -0.03 & -0.00 & -0.00 & -0.02 & 0.12 & 0.01 \\ 
  inflation & 0.00 & 0.00 & -0.01 & -0.04 & -0.03 & 0.02 & 0.04 & -0.02 & 0.02 & -0.04 &  & -0.03 & 0.00 & -0.02 & 0.02 & -0.01 \\ 
  gdp & 0.00 & 0.00 & 0.00 & -0.05 & -0.05 & -0.08 & -0.03 & 0.03 & 0.04 & -0.01 & -0.05 &  & -0.00 & -0.01 & 0.01 & -0.01 \\ 
  midrate & 0.00 & 0.06 & 0.06 & 0.07 & -0.04 & -0.00 & 0.03 & 0.04 & -0.02 & -0.00 & 0.00 & -0.01 &  & -0.08 & -0.00 & 0.02 \\ 
  hash & 0.02 & -0.03 & -0.02 & -0.01 & -0.02 & -0.01 & -0.00 & 0.00 & -0.04 & -0.03 & 0.01 & -0.01 & -0.08 &  & 0.08 & -0.00 \\ 
  tot.btc & -0.00 & -0.06 & -0.07 & -0.02 & 0.02 & 0.02 & 0.02 & -0.01 & 0.04 & 0.11 & -0.00 & -0.01 & -0.02 & 0.32 &  & -0.00 \\ 
  halv & -0.01 & -0.20 & -0.32 & -0.25 & -0.58 & 0.04 & 0.03 & 0.08 & -0.04 & -0.02 & 0.02 & -0.01 & 0.05 & -0.01 & 0.03 &  \\ \bottomrule
\end{tabular}
}
\end{sidewaystable}

\clearpage

\subsection{Additional plots}
\label{appA_figures} 

\begin{figure}[h!]
\centering
\caption{Evolution of the covariates included in the analysis.}
\label{fig:cov_before}
\includegraphics[scale=0.6]{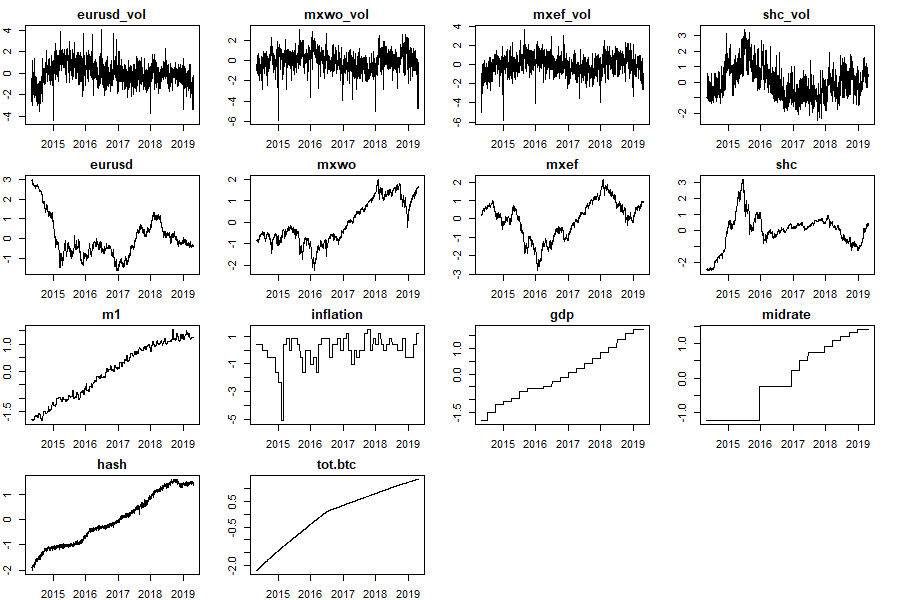}
\caption{Evolution of the covariates included in the analysis, scaled and made stationary.}
\includegraphics[scale=0.6]{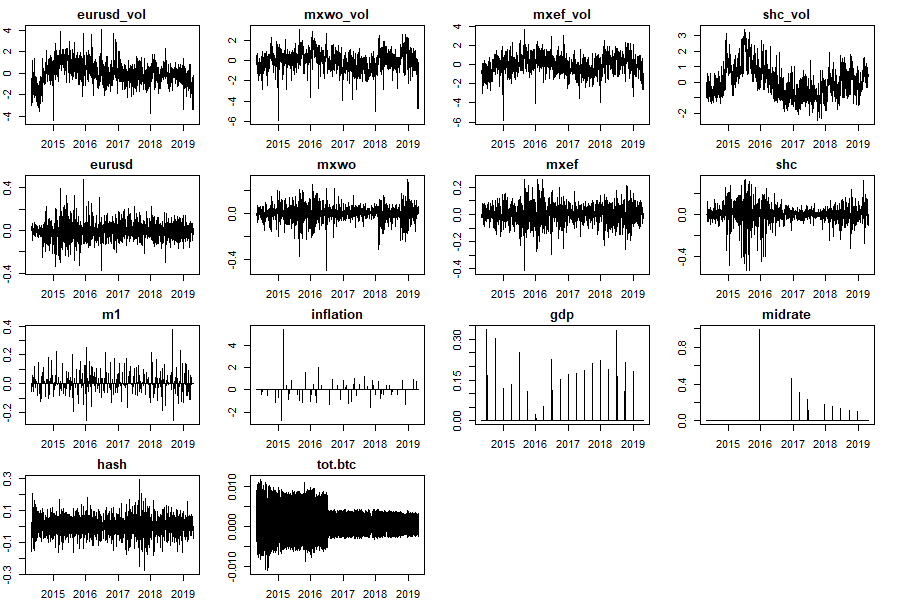}
\end{figure}

\clearpage

\newgeometry{bottom=2cm,top=1.5cm,left=2cm,right=2cm}
\begin{figure}
\caption{Residuals diagnostics (autocorrelation function and Normal QQ Plot) of the six independent C-ARIMA models fitted to the \textbf{Garman-Klass volatility proxy} (in log scale) up to the day before each intervention (four announcements and two futures).}
\label{fig:residuals_GK}
\centering
\begin{tabular}{m{3cm}m{12cm}}
 Announc. 1 & \includegraphics[scale = 0.38]{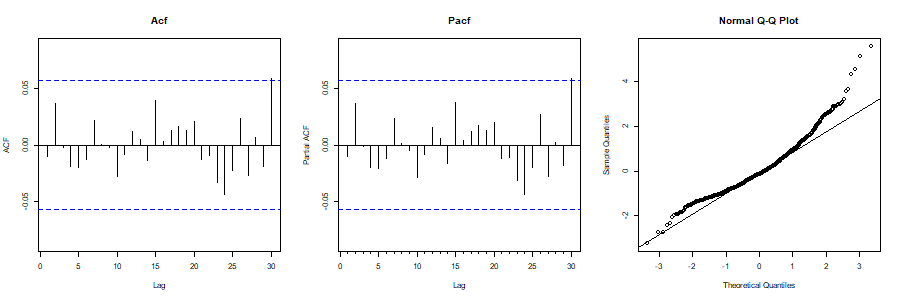} \\
 Announc. 2 & \includegraphics[scale = 0.38]{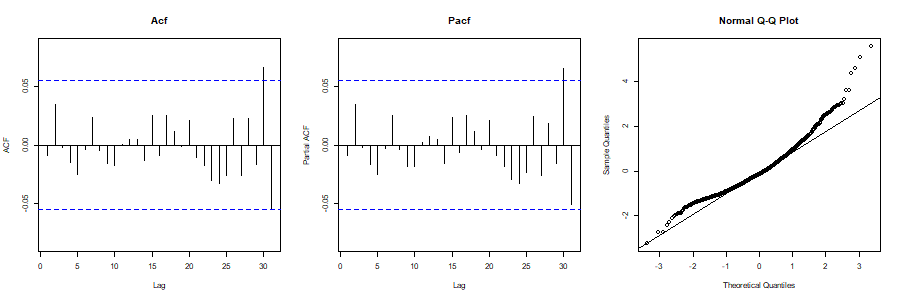} \\
 Announc. 3 & \includegraphics[scale = 0.38]{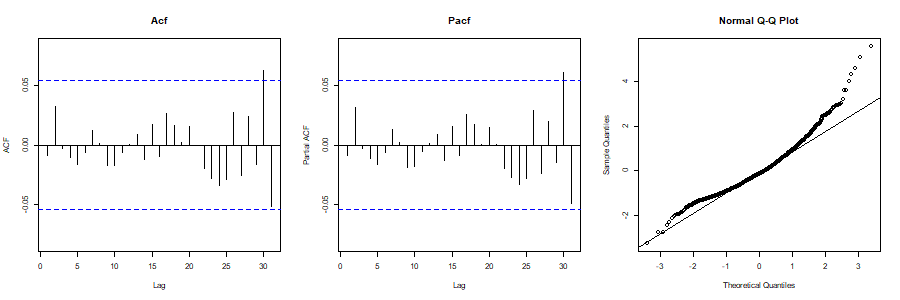} \\
 Announc. 4 & \includegraphics[scale = 0.38]{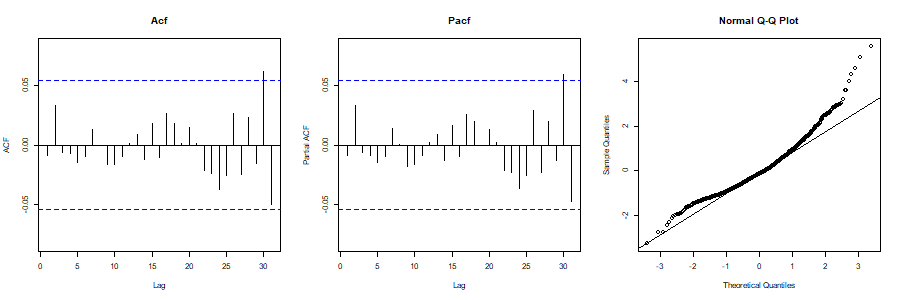} \\
CBOE & \includegraphics[scale = 0.38]{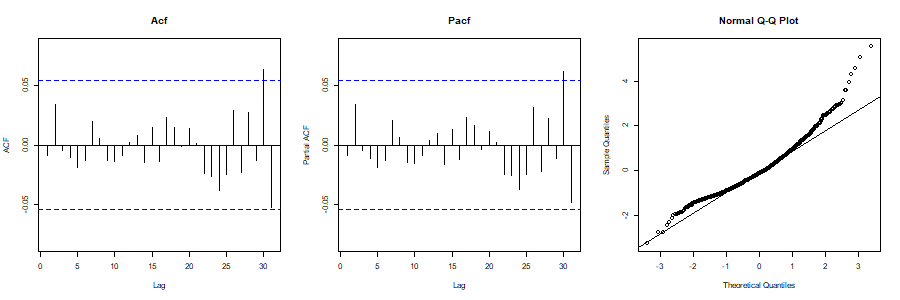} \\
CME & \includegraphics[scale = 0.38]{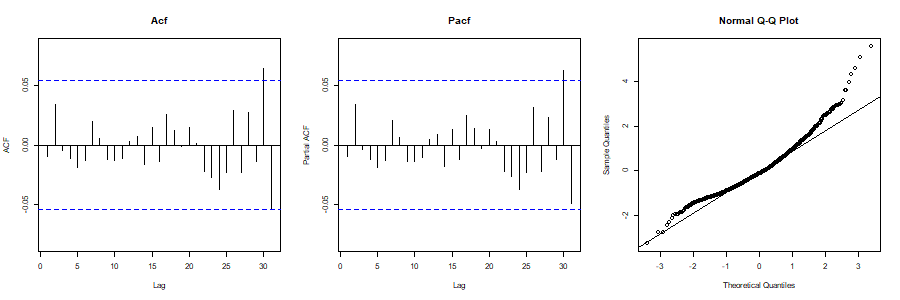} \\
 
\end{tabular}
\end{figure}

\clearpage


\clearpage

%

\begin{figure}
\caption{Residuals diagnostics (autocorrelation function and Normal QQ Plot) of the six independent C-ARIMA models fitted to \textbf{Bitcoin daily volumes} (in log scale) up to the day before each intervention (four announcements and two futures).}
\label{fig:residuals_volumes}
\centering
\begin{tabular}{m{3cm}m{12cm}}
 Announc. 1 & \includegraphics[scale = 0.38]{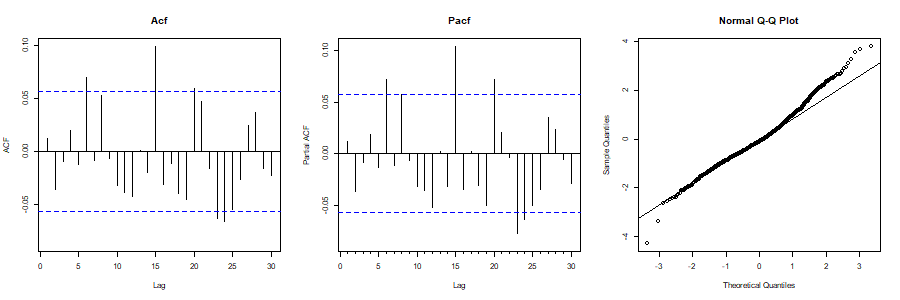} \\
 Announc. 2 & \includegraphics[scale = 0.38]{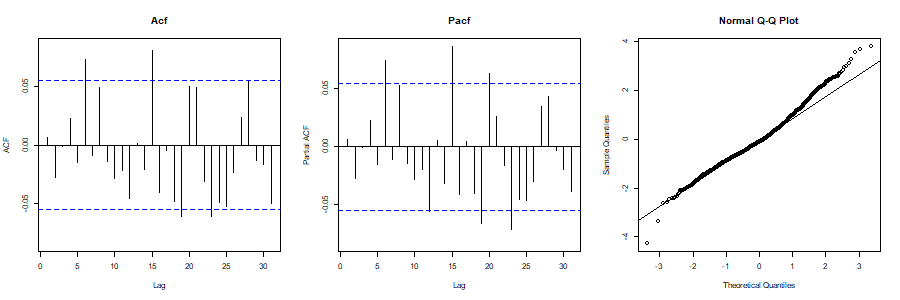} \\
 Announc. 3 & \includegraphics[scale = 0.38]{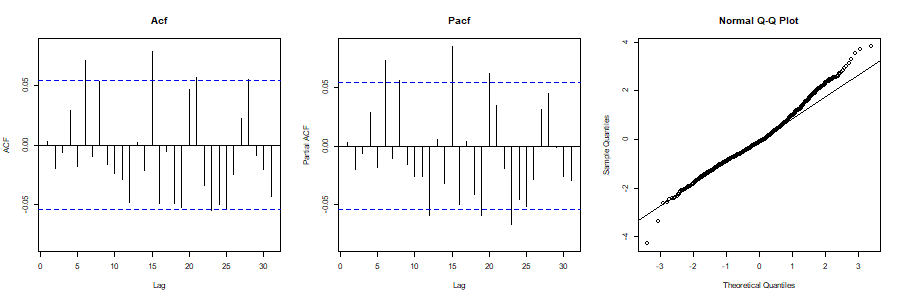} \\
 Announc. 4 & \includegraphics[scale = 0.38]{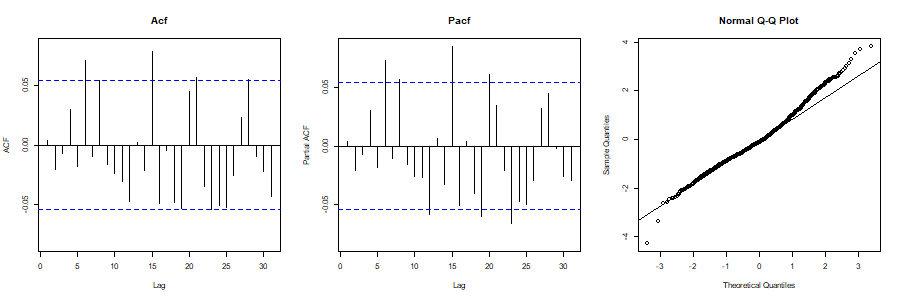} \\
CBOE & \includegraphics[scale = 0.38]{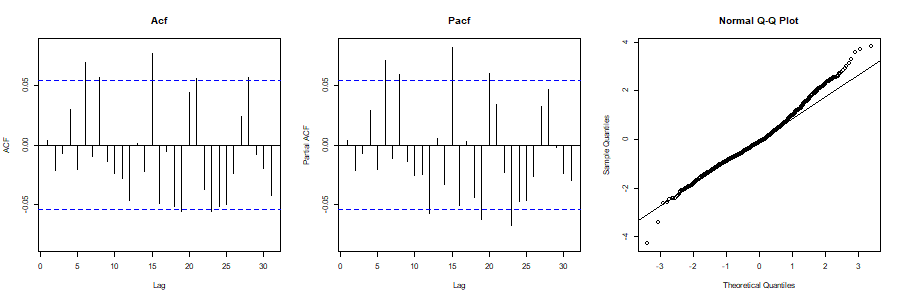} \\
CME & \includegraphics[scale = 0.38]{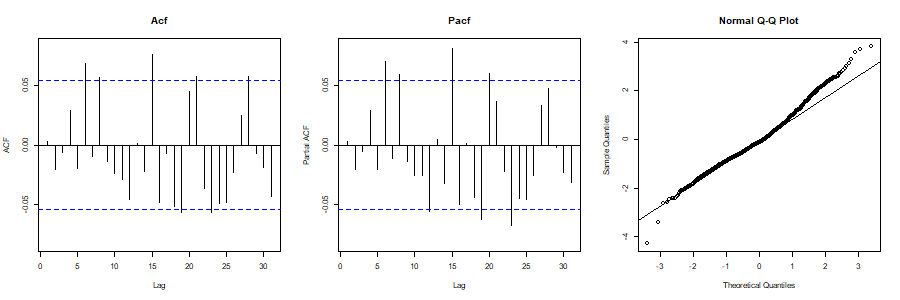} \\
 
\end{tabular}
\end{figure}
\restoregeometry

\clearpage

\end{appendices}

\end{document}

%% file: Tables/BTC_gk_est.tex
\begin{tabular}{@{\extracolsep{5pt}}lD{.}{.}{-3} D{.}{.}{-3} D{.}{.}{-3} D{.}{.}{-3} D{.}{.}{-3} D{.}{.}{-3} } 
\\[-1.8ex]\hline 
\hline \\[-1.8ex] 
\\[-1.8ex] & \multicolumn{1}{c}{Ann.1} & \multicolumn{1}{c}{Ann.2} & \multicolumn{1}{c}{Ann.3} & \multicolumn{1}{c}{Ann.4} & \multicolumn{1}{c}{CBOE} & \multicolumn{1}{c}{CME} \\ 
\hline \\[-1.8ex] 
  $\phi_1$ & 1.257^{***} & 1.244^{***} & 1.251^{***} & 1.248^{***} & 1.254^{***} & 1.255^{***} \\ 
  & (0.059) & (0.058) & (0.057) & (0.057) & (0.056) & (0.056) \\ 
  $\phi_2$ & -0.288^{***} & -0.275^{***} & -0.281^{***} & -0.278^{***} & -0.282^{***} & -0.283^{***} \\ 
  & (0.051) & (0.050) & (0.049) & (0.049) & (0.049) & (0.048) \\ 
  $\theta_1$ & -0.783^{***} & -0.778^{***} & -0.785^{***} & -0.783^{***} & -0.784^{***} & -0.785^{***} \\ 
  & (0.046) & (0.045) & (0.044) & (0.044) & (0.043) & (0.044) \\ 
  $c$ & -3.788^{***} & -3.775^{***} & -3.769^{***} & -3.769^{***} & -3.765^{***} & -3.767^{***} \\ 
  & (0.118) & (0.119) & (0.123) & (0.123) & (0.129) & (0.126) \\ 
  eurusd\_vol & 0.004 & 0.003 & 0.0004 & 0.0002 & -0.002 & -0.003 \\ 
  & (0.021) & (0.020) & (0.020) & (0.020) & (0.020) & (0.020) \\ 
  mxwo\_vol & 0.007 & 0.002 & 0.002 & 0.003 & 0.002 & 0.002 \\ 
  & (0.024) & (0.023) & (0.023) & (0.023) & (0.022) & (0.022) \\ 
  mxef\_vol & 0.002 & 0.005 & 0.006 & 0.006 & 0.009 & 0.009 \\ 
  & (0.021) & (0.020) & (0.020) & (0.020) & (0.020) & (0.020) \\ 
  shc\_vol & -0.013 & -0.021 & -0.021 & -0.021 & -0.017 & -0.017 \\ 
  & (0.028) & (0.027) & (0.027) & (0.027) & (0.027) & (0.027) \\ 
  eurusd & 0.206 & 0.183 & 0.187 & 0.185 & 0.193 & 0.200 \\ 
  & (0.187) & (0.181) & (0.181) & (0.181) & (0.181) & (0.181) \\ 
  mxwo & -0.114 & -0.112 & -0.076 & -0.084 & -0.078 & -0.078 \\ 
  & (0.233) & (0.229) & (0.229) & (0.229) & (0.229) & (0.228) \\ 
  mxef & -0.084 & -0.060 & -0.099 & -0.093 & -0.092 & -0.090 \\ 
  & (0.302) & (0.294) & (0.293) & (0.293) & (0.292) & (0.291) \\ 
  shc & 0.022 & 0.023 & 0.037 & 0.038 & 0.043 & 0.048 \\ 
  & (0.186) & (0.184) & (0.184) & (0.184) & (0.184) & (0.184) \\ 
  m1 & -0.252 & -0.273 & -0.262 & -0.263 & -0.275 & -0.275 \\ 
  & (0.357) & (0.346) & (0.346) & (0.346) & (0.346) & (0.346) \\ 
  inflation & -0.004 & -0.017 & -0.006 & -0.008 & -0.009 & -0.009 \\ 
  & (0.063) & (0.061) & (0.060) & (0.060) & (0.060) & (0.060) \\ 
  gdp & 0.481 & 0.309 & 0.315 & 0.312 & 0.322 & 0.325 \\ 
  & (0.459) & (0.437) & (0.439) & (0.439) & (0.439) & (0.438) \\ 
  midrate & 0.156 & 0.141 & 0.140 & 0.141 & 0.140 & 0.158 \\ 
  & (0.318) & (0.316) & (0.318) & (0.318) & (0.318) & (0.314) \\ 
  hash & 0.035 & -0.046 & -0.046 & -0.042 & -0.041 & -0.036 \\ 
  & (0.172) & (0.156) & (0.153) & (0.153) & (0.152) & (0.152) \\ 
  tot.btc & 0.005 & -0.135 & -0.234 & -0.159 & -0.193 & -0.190 \\ 
  & (2.982) & (2.944) & (2.952) & (2.956) & (2.951) & (2.943) \\ 
  halv & -0.032 & 0.025 & 0.064 & 0.069 & 0.096 & 0.089 \\ 
  & (0.190) & (0.183) & (0.185) & (0.185) & (0.191) & (0.188) \\ 
 \hline \\[-1.8ex] 
$\hat{\tau}^{(n)}$ & -0.55  & -0.06   & 0.39  & -0.33  &  &   \\
$\hat{\bar{\tau}}^{(CBOE)}$ & & & & &  -0.26 & \\
$\hat{\bar{\tau}}^{(CME)}_{t = 7}$ & & & & & & 0.77^{*} \\
$\hat{\bar{\tau}}^{(CME)}_{t = 14}$ & & & & & &  0.73^{*} \\
$\hat{\bar{\tau}}^{CME}_{t = 21}$   & & & & & &  0.70^{.} \\ 
                                    \hline                                                                                                   
Observations & \multicolumn{1}{c}{1,187} & \multicolumn{1}{c}{1,277} & \multicolumn{1}{c}{1,308} & \multicolumn{1}{c}{1,311} & \multicolumn{1}{c}{1,318} & \multicolumn{1}{c}{1,325} \\ 
$\sigma^{2}$ & \multicolumn{1}{c}{0.253} & \multicolumn{1}{c}{0.249} & \multicolumn{1}{c}{0.251} & \multicolumn{1}{c}{0.252} & \multicolumn{1}{c}{0.252} & \multicolumn{1}{c}{0.251} \\ 
Bayesian Inf. Crit. & \multicolumn{1}{c}{1,860.245} & \multicolumn{1}{c}{1,971.905} & \multicolumn{1}{c}{2,030.188} & \multicolumn{1}{c}{2,036.363} & \multicolumn{1}{c}{2,050.867} & \multicolumn{1}{c}{2,055.838} \\ 
\hline
\hline \\[-1.8ex] 
\textit{Note:}  & \multicolumn{6}{r}{$^{\boldsymbol{\cdot}}$p$<$0.1; $^{*}$p$<$0.05; $^{**}$p$<$0.01; $^{***}$p$<$0.001} \\ 
\end{tabular} 

%% file: Tables/BTC_volumes_est.tex
\begin{tabular}{@{\extracolsep{5pt}}lD{.}{.}{-3} D{.}{.}{-3} D{.}{.}{-3} D{.}{.}{-3} D{.}{.}{-3} D{.}{.}{-3} } 
\\[-1.8ex]\hline 
\hline \\[-1.8ex] 
\\[-1.8ex] & \multicolumn{1}{c}{Ann.1} & \multicolumn{1}{c}{Ann.2} & \multicolumn{1}{c}{Ann.3} & \multicolumn{1}{c}{Ann.4} & \multicolumn{1}{c}{CBOE} & \multicolumn{1}{c}{CME} \\ 
\hline \\[-1.8ex] 
   $\phi_1$ & 0.463^{***} & 0.464^{***} & 0.463^{***} & 0.462^{***} & 0.465^{***} & 0.466^{***} \\ 
  & (0.035) & (0.033) & (0.033) & (0.033) & (0.033) & (0.033) \\ 
  $\theta_1$ & -0.919^{***} & -0.921^{***} & -0.920^{***} & -0.920^{***} & -0.921^{***} & -0.921^{***} \\ 
  & (0.017) & (0.017) & (0.017) & (0.017) & (0.017) & (0.017) \\ 
  $\Phi_1$ & 0.146^{***} & 0.144^{***} & 0.137^{***} & 0.137^{***} & 0.140^{***} & 0.139^{***} \\ 
  & (0.030) & (0.029) & (0.029) & (0.029) & (0.029) & (0.029) \\ 
  $\Phi_2$ & 0.095^{**} & 0.097^{***} & 0.097^{***} & 0.097^{***} & 0.094^{***} & 0.093^{***} \\ 
  & (0.030) & (0.029) & (0.028) & (0.028) & (0.028) & (0.028) \\ 
  eurusd\_vol & 0.016 & 0.012 & 0.007 & 0.007 & 0.006 & 0.006 \\ 
  & (0.016) & (0.016) & (0.016) & (0.016) & (0.016) & (0.015) \\ 
  mxwo\_vol & 0.017 & 0.010 & 0.009 & 0.009 & 0.008 & 0.008 \\ 
  & (0.019) & (0.018) & (0.018) & (0.018) & (0.018) & (0.018) \\ 
  mxef\_vol & 0.013 & 0.019 & 0.021 & 0.021 & 0.022 & 0.023 \\ 
  & (0.016) & (0.016) & (0.016) & (0.016) & (0.016) & (0.016) \\ 
  shc\_vol & -0.017 & -0.022 & -0.023 & -0.023 & -0.021 & -0.021 \\ 
  & (0.022) & (0.021) & (0.021) & (0.021) & (0.021) & (0.021) \\ 
  eurusd & 0.076 & 0.082 & 0.079 & 0.079 & 0.079 & 0.081 \\ 
  & (0.143) & (0.140) & (0.139) & (0.139) & (0.139) & (0.138) \\ 
  mxwo & -0.448^{*} & -0.443^{*} & -0.420^{*} & -0.418^{*} & -0.413^{*} & -0.402^{*} \\ 
  & (0.178) & (0.176) & (0.175) & (0.175) & (0.175) & (0.174) \\ 
  mxef & 0.340 & 0.311 & 0.277 & 0.274 & 0.275 & 0.262 \\ 
  & (0.232) & (0.227) & (0.225) & (0.225) & (0.224) & (0.223) \\ 
  shc & -0.105 & -0.088 & -0.085 & -0.085 & -0.083 & -0.082 \\ 
  & (0.142) & (0.141) & (0.141) & (0.141) & (0.141) & (0.141) \\ 
  m1 & 0.155 & 0.124 & 0.125 & 0.125 & 0.118 & 0.118 \\ 
  & (0.267) & (0.259) & (0.259) & (0.258) & (0.258) & (0.258) \\ 
  inflation & 0.041 & 0.036 & 0.033 & 0.034 & 0.032 & 0.033 \\ 
  & (0.048) & (0.047) & (0.046) & (0.046) & (0.046) & (0.046) \\ 
  gdp & 0.302 & 0.256 & 0.269 & 0.269 & 0.272 & 0.275 \\ 
  & (0.353) & (0.337) & (0.338) & (0.337) & (0.337) & (0.336) \\ 
  midrate & 0.253 & 0.248 & 0.255 & 0.255 & 0.252 & 0.257 \\ 
  & (0.245) & (0.245) & (0.245) & (0.245) & (0.245) & (0.241) \\ 
  hash & 0.104 & 0.069 & 0.088 & 0.088 & 0.084 & 0.082 \\ 
  & (0.131) & (0.119) & (0.116) & (0.116) & (0.116) & (0.115) \\ 
  tot.btc & 5.623^{.} & 5.545^{.} & 5.613^{.} & 5.630^{.} & 5.600^{.} & 5.602^{.} \\ 
  & (2.977) & (2.949) & (2.911) & (2.911) & (2.908) & (2.893) \\ 
  halv & -0.337 & -0.351 & -0.354 & -0.354 & -0.352 & -0.355 \\ 
  & (0.287) & (0.284) & (0.283) & (0.283) & (0.284) & (0.283) \\ 
 \hline \\[-1.8ex] 
$\hat{\tau}^{(n)}$ & -0.39  & 0.21   & -0.11  & -0.30  &  &   \\
$\hat{\bar{\tau}}^{(CBOE)}$ & & & & &  -0.20^{**} & \\
$\hat{\bar{\tau}}^{(CME)}_{t = 7}$ & & & & & & 0.33^{***} \\
$\hat{\bar{\tau}}^{(CME)}_{t = 14}$ & & & & & &  0.16^{***} \\
$\hat{\bar{\tau}}^{CME}_{t = 21}$   & & & & & & 0.06^{*} \\ 
\hline                                                                                                  
Observations & \multicolumn{1}{c}{1,186} & \multicolumn{1}{c}{1,276} & \multicolumn{1}{c}{1,307} & \multicolumn{1}{c}{1,310} & \multicolumn{1}{c}{1,317} & \multicolumn{1}{c}{1,324} \\ 
$\sigma^{2}$ & \multicolumn{1}{c}{0.162} & \multicolumn{1}{c}{0.161} & \multicolumn{1}{c}{0.161} & \multicolumn{1}{c}{0.161} & \multicolumn{1}{c}{0.161} & \multicolumn{1}{c}{0.160} \\ 
Bayesian Inf. Crit. & \multicolumn{1}{c}{1,329.856} & \multicolumn{1}{c}{1,417.618} & \multicolumn{1}{c}{1,448.471} & \multicolumn{1}{c}{1,449.489} & \multicolumn{1}{c}{1,457.242} & \multicolumn{1}{c}{1,459.16} \\ 
\hline
\hline \\[-1.8ex] 
\textit{Note:}  & \multicolumn{6}{r}{$^{\boldsymbol{\cdot}}$p$<$0.1; $^{*}$p$<$0.05; $^{**}$p$<$0.01; $^{***}$p$<$0.001} \\ 
\end{tabular} 

%% file: Tables/BTC_mediation.tex
\begin{tabular}{@{\extracolsep{5pt}}lD{.}{.}{-3} D{.}{.}{-3} } 
\\[-1.8ex]\hline 
\hline \\[-1.8ex] 
\\[-1.8ex] & \multicolumn{1}{c}{CBOE} & \multicolumn{1}{c}{CME} \\ 
\hline \\[-1.8ex] 
 $\phi_1$ & 0.936^{***} & 0.934^{***} \\ 
  & (0.014) & (0.014) \\ 
 $\theta_1$ & -0.590^{***} & -0.589^{***} \\ 
  & (0.035) & (0.035) \\ 
 $c$ & -3.785^{***} & -3.786^{***} \\ 
  & (0.109) & (0.107) \\ 
  eurusd\_vol & 0.001 & -0.0001 \\ 
  & (0.020) & (0.020) \\ 
  mxwo\_vol & -0.006 & -0.006 \\ 
  & (0.022) & (0.022) \\ 
  mxef\_vol & 0.007 & 0.007 \\ 
  & (0.020) & (0.020) \\ 
  shc\_vol & -0.007 & -0.007 \\ 
  & (0.026) & (0.026) \\ 
  eurusd & 0.264 & 0.269 \\ 
  & (0.186) & (0.186) \\ 
  mxwo & -0.110 & -0.106 \\ 
  & (0.236) & (0.235) \\ 
  mxef & -0.099 & -0.102 \\ 
  & (0.299) & (0.298) \\ 
  shc & 0.111 & 0.115 \\ 
  & (0.186) & (0.186) \\ 
  m1 & -0.181 & -0.181 \\ 
  & (0.354) & (0.354) \\ 
  inflation & 0.005 & 0.004 \\ 
  & (0.062) & (0.062) \\ 
  gdp & 0.155 & 0.158 \\ 
  & (0.447) & (0.447) \\ 
  midrate & 0.184 & 0.189 \\ 
  & (0.327) & (0.323) \\ 
  hash & -0.003 & 0.0002 \\ 
  & (0.164) & (0.164) \\ 
  tot.btc & 4.140 & 4.129 \\ 
  & (3.260) & (3.252) \\ 
  halv & 0.126 & 0.119 \\ 
  & (0.166) & (0.163) \\ 
  $(1-L) \V_{t-1}$ & 0.189^{***} & 0.190^{***} \\ 
  & (0.031) & (0.031) \\ 
  $(1-L)\V_{t-2}$ & 0.124^{***} & 0.125^{***} \\ 
  & (0.031) & (0.031) \\ 
  $(1-L)\V_{t-3}$ & 0.075^{*} & 0.077^{**} \\ 
  & (0.030) & (0.029) \\ 
 \hline \\[-1.8ex] 
 $\hat{\bar{\tau}}^{(CBOE)}$ & -0.30 & \\
$\hat{\bar{\tau}}^{(CME)}_{t = 7}$ & &  0.83^{*} \\
$\hat{\bar{\tau}}^{(CME)}_{t = 14}$ & &   0.82^{*} \\
$\hat{\bar{\tau}}^{CME}_{t = 21}$   & &   0.79^{*} \\ 
\hline
Observations & \multicolumn{1}{c}{1,314} & \multicolumn{1}{c}{1,321} \\ 
$\sigma^{2}$ & \multicolumn{1}{c}{0.251} & \multicolumn{1}{c}{0.250} \\ 
Bayesian Inf. Crit. & \multicolumn{1}{c}{2,048.938} & \multicolumn{1}{c}{2,053.79} \\ 
\hline  
\hline \\[-1.8ex] 
\textit{Note:}  & \multicolumn{2}{r}{$^{\boldsymbol{\cdot}}$p$<$0.1; $^{*}$p$<$0.05; $^{**}$p$<$0.01; $^{***}$p$<$0.001} \\ 
\end{tabular}

%% file: Tables/BTC_gk_est_app.tex
\begin{tabular}{@{\extracolsep{5pt}}lD{.}{.}{-3} D{.}{.}{-3} D{.}{.}{-3} D{.}{.}{-3} D{.}{.}{-3} D{.}{.}{-3} } 
\\[-1.8ex]\hline 
\hline \\[-1.8ex] 
\\[-1.8ex] & \multicolumn{1}{c}{Ann.1} & \multicolumn{1}{c}{Ann.2} & \multicolumn{1}{c}{Ann.3} & \multicolumn{1}{c}{Ann.4} & \multicolumn{1}{c}{CBOE} & \multicolumn{1}{c}{CME} \\ 
\hline \\[-1.8ex] 
 $\phi_1$ & 1.247^{***} & 1.237^{***} & 1.246^{***} & 1.243^{***} & 1.250^{***} & 1.251^{***} \\ 
  & (0.059) & (0.057) & (0.056) & (0.056) & (0.055) & (0.055) \\ 
  $\phi_2$ & -0.280^{***} & -0.268^{***} & -0.276^{***} & -0.273^{***} & -0.278^{***} & -0.280^{***} \\ 
  & (0.051) & (0.049) & (0.048) & (0.049) & (0.048) & (0.048) \\ 
  $\theta_1$ & -0.778^{***} & -0.776^{***} & -0.784^{***} & -0.782^{***} & -0.784^{***} & -0.786^{***} \\ 
  & (0.046) & (0.045) & (0.044) & (0.044) & (0.043) & (0.043) \\ 
  $c$ & -3.800^{***} & -3.770^{***} & -3.745^{***} & -3.743^{***} & -3.731^{***} & -3.735^{***} \\ 
  & (0.099) & (0.097) & (0.100) & (0.100) & (0.105) & (0.103) \\  
 \hline \\[-1.8ex] 
 $\hat{\tau}^{(n)}$ & -0.53   &  -0.04  & 0.40    &  -0.33  &  &   \\
$\hat{\bar{\tau}}^{(CBOE)}$ & & & & &  -0.25 & \\
$\hat{\bar{\tau}}^{(CME)}_{t = 7}$ & & & & & & 0.78^{*} \\
$\hat{\bar{\tau}}^{(CME)}_{t = 14}$ & & & & & & 0.76^{*}   \\
$\hat{\bar{\tau}}^{CME}_{t = 21}$   & & & & & &  0.72^{.} \\ 
\hline                                                                                                   
Observations & \multicolumn{1}{c}{1,187} & \multicolumn{1}{c}{1,277} & \multicolumn{1}{c}{1,308} & \multicolumn{1}{c}{1,311} & \multicolumn{1}{c}{1,318} & \multicolumn{1}{c}{1,325} \\ 
$\sigma^{2}$ & \multicolumn{1}{c}{0.251} & \multicolumn{1}{c}{0.247} & \multicolumn{1}{c}{0.249} & \multicolumn{1}{c}{0.249} & \multicolumn{1}{c}{0.250} & \multicolumn{1}{c}{0.249} \\ 
Bayesian Inf. Crit. & \multicolumn{1}{c}{1,758.394} & \multicolumn{1}{c}{1,868.65} & \multicolumn{1}{c}{1,926.575} & \multicolumn{1}{c}{1,932.734} & \multicolumn{1}{c}{1,947.174} & \multicolumn{1}{c}{1,952.196} \\ 
\hline  
\hline \\[-1.8ex] 
\textit{Note:}  & \multicolumn{6}{r}{$^{\boldsymbol{\cdot}}$p$<$0.1; $^{*}$p$<$0.05; $^{**}$p$<$0.01; $^{***}$p$<$0.001} \\ 
\end{tabular} 

%% file: Tables/BTC_volumes_est_app.tex
\begin{tabular}{@{\extracolsep{5pt}}lD{.}{.}{-3} D{.}{.}{-3} D{.}{.}{-3} D{.}{.}{-3} D{.}{.}{-3} D{.}{.}{-3} } 
\\[-1.8ex]\hline 
\hline \\[-1.8ex] 
\\[-1.8ex] & \multicolumn{1}{c}{Ann.1} & \multicolumn{1}{c}{Ann.2} & \multicolumn{1}{c}{Ann.3} & \multicolumn{1}{c}{Ann.4} & \multicolumn{1}{c}{CBOE} & \multicolumn{1}{c}{CME} \\ 
\hline \\[-1.8ex] 
   $\phi_1$ & 0.458^{***} & 0.469^{***} & 0.470^{***} & 0.469^{***} & 0.472^{***} & 0.474^{***} \\ 
  & (0.046) & (0.044) & (0.043) & (0.043) & (0.043) & (0.042) \\ 
  $\theta_1$ & -0.927^{***} & -0.937^{***} & -0.937^{***} & -0.937^{***} & -0.937^{***} & -0.939^{***} \\ 
  & (0.027) & (0.025) & (0.025) & (0.026) & (0.025) & (0.024) \\ 
  $\Phi_1$ & 0.134^{***} & 0.135^{***} & 0.127^{***} & 0.127^{***} & 0.132^{***} & 0.131^{***} \\ 
  & (0.035) & (0.034) & (0.033) & (0.033) & (0.033) & (0.033) \\ 
  $\Phi_2$ & 0.124^{***} & 0.126^{***} & 0.123^{***} & 0.123^{***} & 0.119^{***} & 0.118^{***} \\ 
  & (0.034) & (0.032) & (0.032) & (0.032) & (0.032) & (0.032) \\ 
  eurusd\_vol & 0.014 & 0.010 & 0.004 & 0.005 & 0.003 & 0.002 \\ 
  & (0.019) & (0.018) & (0.018) & (0.018) & (0.018) & (0.017) \\ 
  mxwo\_vol & 0.007 & 0.0002 & -0.001 & -0.001 & -0.003 & -0.002 \\ 
  & (0.021) & (0.020) & (0.020) & (0.020) & (0.020) & (0.020) \\ 
  mxef\_vol & 0.018 & 0.026 & 0.028 & 0.028 & 0.030^{.} & 0.031^{.} \\ 
  & (0.019) & (0.018) & (0.018) & (0.018) & (0.018) & (0.018) \\ 
  shc\_vol & 0.003 & -0.006 & -0.007 & -0.007 & -0.005 & -0.006 \\ 
  & (0.024) & (0.023) & (0.023) & (0.023) & (0.023) & (0.022) \\ 
  eurusd & 0.242 & 0.236 & 0.233 & 0.233 & 0.233 & 0.234 \\ 
  & (0.155) & (0.150) & (0.149) & (0.149) & (0.149) & (0.148) \\ 
  mxwo & -0.603^{**} & -0.595^{**} & -0.566^{**} & -0.564^{**} & -0.558^{**} & -0.542^{**} \\ 
  & (0.197) & (0.194) & (0.193) & (0.192) & (0.192) & (0.191) \\ 
  mxef & 0.422^{.} & 0.390 & 0.347 & 0.345 & 0.345 & 0.327 \\ 
  & (0.254) & (0.247) & (0.245) & (0.245) & (0.243) & (0.242) \\ 
  shc & -0.071 & -0.055 & -0.050 & -0.050 & -0.048 & -0.048 \\ 
  & (0.155) & (0.154) & (0.153) & (0.153) & (0.153) & (0.153) \\ 
  m1 & 0.121 & 0.096 & 0.100 & 0.100 & 0.090 & 0.091 \\ 
  & (0.292) & (0.281) & (0.280) & (0.280) & (0.280) & (0.279) \\ 
  inflation & 0.032 & 0.026 & 0.024 & 0.024 & 0.023 & 0.023 \\ 
  & (0.049) & (0.048) & (0.047) & (0.047) & (0.047) & (0.047) \\ 
  gdp & 0.630 & 0.523 & 0.539 & 0.539 & 0.541 & 0.544 \\ 
  & (0.474) & (0.439) & (0.439) & (0.439) & (0.439) & (0.438) \\ 
  midrate & 0.267 & 0.257 & 0.266 & 0.266 & 0.262 & 0.263 \\ 
  & (0.244) & (0.243) & (0.243) & (0.243) & (0.243) & (0.239) \\ 
  hash & 0.117 & 0.059 & 0.085 & 0.085 & 0.079 & 0.076 \\ 
  & (0.154) & (0.134) & (0.130) & (0.130) & (0.129) & (0.129) \\ 
  tot.btc & 7.210^{.} & 7.233^{.} & 7.397^{*} & 7.430^{*} & 7.404^{*} & 7.401^{*} \\ 
  & (3.832) & (3.780) & (3.711) & (3.710) & (3.704) & (3.680) \\ 
  halv & -0.369 & -0.406 & -0.412 & -0.412 & -0.409 & -0.416 \\ 
  & (0.281) & (0.273) & (0.272) & (0.272) & (0.273) & (0.270) \\ 
 \hline \\[-1.8ex] 
$\hat{\tau}^{(n)}$ & -0.41  & 0.22   & -0.09  & -0.30  &  &   \\
$\hat{\bar{\tau}}^{(CBOE)}$ & & & & &  -0.15^{*} & \\
$\hat{\bar{\tau}}^{(CME)}_{t = 7}$ & & & & & & 0.33^{***} \\
$\hat{\bar{\tau}}^{(CME)}_{t = 14}$ & & & & & &  0.16^{***} \\
$\hat{\bar{\tau}}^{CME}_{t = 21}$   & & & & & & 0.06^{*} \\ 
\hline                                                                                                  
Observations & \multicolumn{1}{c}{912} & \multicolumn{1}{c}{1,002} & \multicolumn{1}{c}{1,033} & \multicolumn{1}{c}{1,036} & \multicolumn{1}{c}{1,043} & \multicolumn{1}{c}{1,050} \\ 
$\sigma^{2}$ & \multicolumn{1}{c}{0.159} & \multicolumn{1}{c}{0.158} & \multicolumn{1}{c}{0.158} & \multicolumn{1}{c}{0.157} & \multicolumn{1}{c}{0.158} & \multicolumn{1}{c}{0.157} \\ 
Bayesian Inf. Crit. & \multicolumn{1}{c}{1,026.865} & \multicolumn{1}{c}{1,114.535} & \multicolumn{1}{c}{1,145.139} & \multicolumn{1}{c}{1,146.05} & \multicolumn{1}{c}{1,153.949} & \multicolumn{1}{c}{1,155.693} \\ 
\hline 
\hline \\[-1.8ex] 
\textit{Note:}  & \multicolumn{6}{r}{$^{\boldsymbol{\cdot}}$p$<$0.1; $^{*}$p$<$0.05; $^{**}$p$<$0.01; $^{***}$p$<$0.001} \\ 
\end{tabular} 

%% file: Tables/BTC_mediation_volatility.tex
\begin{tabular}{@{\extracolsep{5pt}}lD{.}{.}{-3} D{.}{.}{-3} } 
\\[-1.8ex]\hline 
\hline \\[-1.8ex] 
\\[-1.8ex] & \multicolumn{1}{c}{CBOE} & \multicolumn{1}{c}{CME} \\ 
\\[-1.8ex] & \multicolumn{1}{c}{(1)} & \multicolumn{1}{c}{(2)}\\ 
\hline \\[-1.8ex] 
  $\phi_1$ & 0.474^{***} & 0.474^{***} \\ 
  & (0.046) & (0.046) \\ 
  $\theta_1$ & -0.914^{***} & -0.915^{***} \\ 
  & (0.018) & (0.018) \\ 
  $\Phi_1$ & 0.151^{***} & 0.150^{***} \\ 
  & (0.029) & (0.029) \\ 
  eurusd\_vol & 0.005 & 0.004 \\ 
  & (0.016) & (0.016) \\ 
  mxwo\_vol & 0.010 & 0.011 \\ 
  & (0.018) & (0.018) \\ 
  mxef\_vol & 0.024 & 0.025 \\ 
  & (0.016) & (0.016) \\ 
  shc\_vol & -0.023 & -0.023 \\ 
  & (0.021) & (0.021) \\ 
  eurusd & 0.074 & 0.076 \\ 
  & (0.140) & (0.140) \\ 
  mxwo & -0.363^{*} & -0.354^{*} \\ 
  & (0.174) & (0.173) \\ 
  mxef & 0.219 & 0.211 \\ 
  & (0.224) & (0.223) \\ 
  shc & -0.061 & -0.060 \\ 
  & (0.143) & (0.142) \\ 
  m1 & 0.140 & 0.140 \\ 
  & (0.262) & (0.262) \\ 
  inflation & 0.037 & 0.037 \\ 
  & (0.046) & (0.046) \\ 
  gdp & 0.301 & 0.304 \\ 
  & (0.337) & (0.337) \\ 
  midrate & 0.232 & 0.243 \\ 
  & (0.245) & (0.242) \\ 
  hash & 0.080 & 0.080 \\ 
  & (0.116) & (0.116) \\ 
  tot.btc & 5.679^{*} & 5.693^{*} \\ 
  & (2.643) & (2.633) \\ 
  halv & -0.386 & -0.390 \\ 
  & (0.292) & (0.291) \\ 
  $Y_{t-1}$ & -0.013 & -0.013 \\ 
  & (0.032) & (0.032) \\ 
  $Y_{t-2}$ & -0.023 & -0.022 \\ 
  & (0.026) & (0.026) \\ 
  $Y_{t-3}$ & -0.013 & -0.012 \\ 
  & (0.023) & (0.023) \\ 
 \hline \\[-1.8ex] 
Observations & \multicolumn{1}{c}{1,314} & \multicolumn{1}{c}{1,321} \\ 
$\sigma^{2}$ & \multicolumn{1}{c}{0.163} & \multicolumn{1}{c}{0.162} \\ 
Bayesian Inf. Crit. & \multicolumn{1}{c}{1,481.343} & \multicolumn{1}{c}{1,483.269} \\ 
\hline  
\hline \\[-1.8ex] 
\textit{Note:}  & \multicolumn{2}{r}{$^{\boldsymbol{\cdot}}$p$<$0.1; $^{*}$p$<$0.05; $^{**}$p$<$0.01; $^{***}$p$<$0.001} \\ 
\end{tabular} 

%% file: Tables/BTC_norm.tex
\begin{tabular}{l D{.}{.}{-2} r D{.}{.}{-2} r D{.}{.}{-2}} 
\\[-1.8ex]\hline 
\hline \\[-1.8ex] 
Item  &  (1)   &  & (2)  &  & (3)  \\ \hline \\[-1.8ex] 
 $\hat{\tau}^{(1)}$          &  -0.55    &  &   -0.39      &  &  \\
                             & (0.50)    &  &  (0.40)      &  & \\[0.3cm]
 $\hat{\tau}^{(2)}$          &  -0.06    &  &   0.21       &  &    \\
                             & (0.50)    &  &  (0.40)      &  & \\[0.3cm]
 $\hat{\tau}^{(3)}$          &  0.39     &  &   -0.11      &  &  \\
                             & (0.50)    &  &  (0.40)      &  & \\[0.3cm]
 $\hat{\tau}^{(4)}$          &  -0.33    &  &   -0.30      &  &  \\
                             & (0.50)    &  &  (0.40)      &  & \\[0.3cm]
 $\hat{\tau}^{(CBOE)}$       &  -0.26    &  &   -0.20^{**} &  & -0.30    \\
                             & (0.38)    &  &  (0.06)      &  & (0.38) \\[0.3cm]
 $\hat{\tau}^{(CME)}_{t=7}$  &  0.77^{*} &  &   0.33^{***} &  & 0.83^{*}  \\
                             & (0.38)    &  &  (0.06)      &  & (0.38)\\[0.3cm]
 $\hat{\tau}^{(CME)}_{t=14}$ &  0.73^{.} &  &   0.16^{***} &  & 0.82^{*}  \\
                             & (0.38)    &  &  (0.04)      &  & (0.39)\\[0.3cm]
 $\hat{\tau}^{(CME)}_{t=21}$ &  0.70^{.} &  &   0.06^{*}   &  & 0.79^{*}  \\
                             & (0.38)    &  &  (0.03)      &  & (0.39) \\[0.3cm]
\hline
\hline \\[-1.8ex]
\textit{Note:}      & \multicolumn{5}{r}{$^{\boldsymbol{\cdot}}$p$<$0.1; $^{*}$p$<$0.05; $^{**}$p$<$0.01; $^{***}$p$<$0.001} \\                  
\end{tabular}